\documentclass[pra,reprint,preprintnumbers,
amsmath,amssymb,
aps, showkeys
]{revtex4-2}

\usepackage{graphicx}
\usepackage{dcolumn}
\usepackage{bm}
\usepackage{amssymb}
\usepackage{amsmath}
\usepackage{color}
\usepackage{soul}
\usepackage{subfig}
\usepackage{appendix}
\usepackage{booktabs}
\usepackage{comment}
\usepackage[hyperindex,breaklinks,hidelinks]{hyperref}

\newcommand{\id}{1\!\!1}
\newcommand{\nn}{\nonumber}

\begin{document}
	
	\title{Critical spectrum of the anisotropic two-photon quantum Rabi model}
	
	\author{Jiong Li$^{1}$}
	\author{Daniel Braak$^{2,}$}
	\email{daniel.braak@uni-a.de}
	\author{Qing-Hu Chen$^{1,3,}$}
	\email{qhchen@zju.edu.cn}
	
	\affiliation{$^{1}$Zhejiang Key Laboratory of Micro-Nano Quantum Chips and Quantum Control, School of Physics, Zhejiang University, Hangzhou 310027, China\\
	$^{2}$Center for Electronic Correlations and Magnetism, Institute of Physics, University of Augsburg, D-86135 Augsburg, Germany\\
	$^{3}$Collaborative Innovation Center of Advanced Microstructures, Nanjing University, Nanjing 210093, China}

	\date{\today}
	
	\begin{abstract}
		The anisotropic two-photon quantum Rabi model is studied using the Bogoliubov operator approach. The doubly degenerate exceptional states are identified through analytical methods. By adjusting the position  of the last exceptional point belonging to two adjacent energy levels, we derive  a condition for the absence of the discrete spectrum at the critical coupling where ``spectral collapse" occurs. In this special case, the spectrum becomes fully continuous above a threshold energy, with no bound states existing, whereas the ground state remains gapped in the general case. This also signals  a quantum phase transition. More  interestingly, we rigorously find  that a finite number of bound states exist between two anisotropy dependent  critical atomic frequencies, with infinitely many bound states beyond this  frequency regime.  In this manner, all issues in the two-photon quantum Rabi model are resolved.
	\end{abstract}
	
	
	\maketitle
	
	\section{Introduction}
	A fundamental issue in quantum optics is the interaction between radiation and matter. The quantum Rabi model (QRM) is recognized as the simplest model describing the light-matter interaction between a qubit (two-level system) and a single quantized bosonic mode \cite{rabi_process_1936}. It finds broad application across various physical domains, such as cavity and circuit quantum electrodynamics (QED) systems, solid state semiconductor systems, trapped ions, and quantum dots \cite{Scully_Zubairy_1997, englund_controlling_2007, hennessy_quantum_2007, niemczyk_circuit_2010, del_valle_two-photon_2010, braak_semi-classical_2016, forn-diaz_ultrastrong_2017, forn-diaz_ultrastrong_2019}. Furthermore, considerable experimental and theoretical efforts have been devoted into generalizations of the QRM, including two-photon interactions \cite{peng_influence_1993, bertet_generating_2002, chen_exact_2012, travenec_solvability_2012, lu_quantum_2017, maciejewski_novel_2017, cong_polaron_2019, xie_generalized_2019, yan_analytic_2020, braak_spectral_2023}, anisotropic couplings \cite{xie_anisotropic_2014, shen_quantum_2017, chen_multiple_2021, hu_excited-state_2023}, Stark-like nonlinear interaction \cite{eckle_generalization_2017, xie_quantum_2019, li_solving_2023, braak2024c} and multimode extensions \cite{zhang_solvability_2013, duan_solution_2015, zhang_hidden_2016, xie_exact_2020, yan_analytical_2022}.
	
	The two-photon Rabi model (tpQRM) has garnered significant attention in recent years, because it may be physically implemented on different platforms \cite{felicetti_spectral_2015, felicetti_two-photon_2018}, has applications in quantum metrology \cite{ying2022}, and is interesting also from the viewpoint of pure mathematics \cite{nakahama2024, hiroshima2024a}. It features a counter-intuitive phenomenon known as ``spectral collapse", first proposed in \cite{ng_exact_nodate}. Spectral collapse occurs when the light-matter coupling approaches a critical value, resulting in the coexistence of discrete and continuous parts within the spectrum, which is discrete below the critical coupling, meaning the spectrum there is of pure-point type \cite{duan_two-photon_2016, chan_bound_2020, lo_manipulating_2020, rico_spectral_2020, braak_spectral_2023}. Although a continuous spectrum does not occur in a confined system (the confinement would require the inclusion of higher non-linear terms in the Hamiltonian) experimental evidence supports the existence of very dense energy levels at the critical point in some experimental realizations \cite{felicetti_spectral_2015, felicetti_two-photon_2018, felicetti_ultrastrong-coupling_2018, piccione_two-photon-interaction_2022}. The presence of counter-rotating terms in the Hamiltonian is vital for the collapse phenomenon \cite{ng_exact_nodate}, so it is illuminating to study a generalization of the tpQRM, the anisotropic two-photon Rabi (atpQRM), with different coupling strengths for the rotating and counter-rotating terms \cite{cui_exact_2017, lo_spectral_2021}. It was found in \cite{lo_spectral_2021} that the atpQRM also exhibits an ``incomplete" spectral collapse like the isotropic model: besides the continuum, there is a discrete set of bound states, similar to the hydrogen atom \cite{duan_two-photon_2016} The discrete part of the spectrum can be derived via a mapping to the problem of a particle with position-dependent effective mass in a finite potential well \cite{lo_spectral_2021}. However, this reduced problem with just a single continuous degree of freedom (the atpQRM has two: the radiation mode and the two-level system) has no known analytical solution. In the following, we shall not adopt the notion ``incomplete" or ``complete" collapse, because it suggests the appearance of an infinitely degenerate state at the collapse point which is actually not present.
	
	In the tpQRM, the discrete bound states vanish at the collapse point when the energy level difference of the qubit turns zero, simply because the qubit is no longer coupled in the spin-boson representation. We have found that the atpQRM can also exhibit this phenomenon under specific, non-trivial coupling conditions. The $G$-function, initially introduced in \cite{braak_integrability_2011} through the Bargmann space representation and later reproduced using the Bogoliubov operator approach (BOA) \cite{chen_exact_2012}, has proven to be a superior tool for analyzing both the QRM and its generalizations. The $G$-function is the only method to obtain the qualitative features of the spectrum analytically via its pole structure, which inspires us to identify the conditions for the presence or absence of discrete bound states by studying the exceptional spectrum of the atpQRM below but close to the critical point.
	
	The paper is organized as follows: Section II presents the $G$-function of the atpQRM which yields the analytical derivation of the exact eigenenergies and the critical (collapse) points. Section III contains the analytical derivation of the doubly degenerate eigenvalues forming part of the exceptional spectrum using the $G$-function and establishes the relationship between these points and the collapse point itself. Sections IV and V investigate the properties of eigenfunctions right at the collapse point for the critical qubit splitting and the general case, respectively. The final section provides a summary of our findings.
	
	\section{Solution with Bogoliubov transformation}
	
	The Hamiltonian of the anisotropic two-photon Rabi model is given by
	\begin{eqnarray}
		H &=& \frac{\Delta}{2} \sigma_{z} + \omega a^{\dagger}a + g_{1} \left[ (a^{\dagger})^{2} \sigma_{-} + \sigma_{+} a^{2} \right] \nonumber \\
		&&+ g_{2} \left[ (a^{\dagger})^{2} \sigma_{+} + \sigma_{-} a^{2} \right]
		\label{H_orgin}
	\end{eqnarray}
	where $a$ and $a^{\dagger}$ are the photon annihilation and creation operators of a single cavity mode with frequency $\omega$, $\Delta $ represents the tunneling matrix element, $g_{1}$ and $g_{2}$ are the qubit-cavity coupling constants for the rotating and count-rotating terms respectively, and $\{ \sigma_{i} \}$ are the Pauli matrices. For simplicity, we set $\omega=1$, $g_{1}=g$, $g_{2}=rg$ throughout this paper. By defining the symmetry operator $\Pi = \sigma_{z} \otimes \exp \left[i \frac{\pi}{2} a^{\dagger}a \right]$ with $\Pi^4=\id$, we find $\Pi H \Pi^{\dagger} = H$, indicating that the Hamiltonian possesses $\mathbb{Z}_{4}$ symmetry.
	
	To proceed, we apply a Bogoliubov (squeezing) transformation $S(\theta)$, followed by a similarity transformation $P$, defined as
	\begin{equation*}
		S(\theta) = e^{ \frac{\theta}{2} \left[(a^{\dagger})^{2}-a^{2} \right]}, \quad
		P = \frac{1}{\sqrt{2}}
		\begin{bmatrix}
			\sqrt{r_2/r_1} & 1 \\ -\sqrt{r_2/r_1} & 1
		\end{bmatrix},
	\end{equation*}
	where $r_{1} = r \sinh^{2} \theta + \cosh^{2} \theta$ and $r_{2} = r \cosh^{2} \theta + \sinh^{2} \theta$.
	The transformed Hamiltonian is given by $H_{s} = S(\theta) P H P^{-1} S(-\theta)$. The parameter $\theta$ of the squeezing transformation is chosen as
	\begin{eqnarray}
		\cosh \theta = \sqrt{ \frac{\beta_{+} + \beta_{-}}{2\beta_{+}}},
		\beta_{\pm} = \sqrt{1 - g^{2} (r \pm 1)^{2}}.
	\end{eqnarray}
	A real $\theta$ is required to render the transformation unitary, which imposes the condition
	\begin{equation}
		g < g_c = \frac{1}{1+r}.
	\end{equation}
	As in the isotropic case, the model changes spectral characteristics at the \emph{critical coupling} $g_c=1/(1+r)$, which depends on the anisotropy $r$, but not on the Bargmann index $q$.
	
	A well-known representation of the Lie algebra $sl_2(\mathbb{R})$ is
	\begin{eqnarray}
		K_{0} = \frac{1}{2} \left( a^{\dagger}a + \frac{1}{2} \right), K_{+} = \frac{(a^{\dagger})^{2}}{2}, K_{-} = \frac{a^{2}}{2}, \nonumber
	\end{eqnarray}
	where
	\begin{equation}
		\left[ K_{0}, K_{\pm} \right] = \pm K_{\pm}, \left[ K_{+}, K_{-}\right] = -2K_{0}.
	\end{equation}
	The Hamiltonian \eqref{H_orgin} can be written in terms of these generators of $sl_2(\mathbb{R})$ and the Hilbert space $L^2(\mathbb{R})$ separates into two irreducible representations of $sl_2(\mathbb{R})$, consisting of vectors $\{(a^\dagger)^n|0\rangle\}$ for even respectively odd $n$.
	They are characterized by the Bargmann index $q$, i.e. the eigenvalue of $K_{0}$ acting on the vacuum (lowest weight) state $ K_0\vert q, 0 \rangle = q \vert q, 0 \rangle$ with $K_-|q,0\rangle=0$. For the even subspace $\mathcal{H}_{\frac{1}{4}} = \left\{\left( a^{\dagger} \right)^{n} \left\vert 0 \right\rangle, n=0, 2, 4...... \right\}$, $q = 1/4$ and for the odd subspace $\mathcal{H}_{\frac{3}{4}} = \left\{ \left( a^{\dagger} \right)^{n} \left\vert 0 \right\rangle, n=1, 3, 5...... \right\}$, $q = 3/4$.
	\begin{eqnarray}
		\left\vert q, n \right\rangle &=& \left\vert 2 \left( q+n-\frac{1}{4} \right) \right\rangle = \frac{\left( a^{\dagger} \right)^{2 \left( q+n-\frac{1}{4} \right)}}{\sqrt{ \left[ 2 \left( q+n-\frac{1}{4} \right)\right] !}} \left\vert 0\right\rangle,  \nonumber \\
		K_{0} \left\vert q, n \right\rangle &=& (q+n) \left\vert q, n \right\rangle,
	\end{eqnarray}
	In terms of $\{ K_{0}, K_{\pm} \}$, the elements of the Hamiltonian $H_{s}$ are expressed as
	\begin{eqnarray}
		H_{11} &=& \beta_{+} \left[ 2 \beta_{-} K_{0} + \frac{g (1-r)^{2} }{\sqrt{r}} K_{+} \right] - \frac{1}{2}, \nonumber \\
		H_{12} &=& - \frac{\Delta}{2} + (1-r^{2}) \left( 2g^{2} K_{0} - \frac{g}{\sqrt{r}} \beta_{-} K_{+} \right), \nonumber \\
		H_{21} &=& - \frac{\Delta}{2} - (1-r^{2}) \left( 2g^{2} K_{0} - \frac{g}{\sqrt{r}} \beta_{-} K_{+} \right), \nonumber \\
		H_{22} &=& \frac{\beta_{-}}{\beta_{+}} \left[ 2 \left( 2 - \beta_{+}^{2} \right) K_{0} - \frac{ g(1+r)^{2} }{\sqrt{r}} \beta_{-} K_{+} - \frac{ 4\sqrt{r}g}{\beta_{-}} K_{-} \right] -\frac{1}{2}. \nonumber
	\end{eqnarray}
	
	The eigenfunctions of $H_{s}$ may be expanded as
	\begin{equation}
		\left\vert \psi^{(q)} \right\rangle =
		\begin{bmatrix}
			\sum_{n=0}^{+\infty }\sqrt{\left[ 2\left( n+q-\frac{1}{4}\right) \right]!} e_{n}^{(q)} \left\vert q,n\right\rangle \\
			\sum_{n=0}^{+\infty }\sqrt{\left[ 2\left( n+q-\frac{1}{4}\right) \right] !} f_{n}^{(q)} \left\vert q,n \right\rangle
		\end{bmatrix}, \label{psi_A_q}
	\end{equation}
	where $e_{n}^{(q)}$ and $f_{n}^{(q)}$ are the expansion coefficients. Using the Schr\"{o}dinger equation $H_{s} \left\vert \psi^{(q)} \right\rangle = E \left\vert \psi^{(q)} \right\rangle $ and projecting onto $\vert q,n \rangle$, we derive a recursive relation for $e_{n}^{(q)}$ and $f_{n}^{(q)}$ as follows:
	\begin{subequations}
		\label{recur_tpQRM}
		\begin{eqnarray}
			e_{n}^{(q)} &=& \frac{g(1-r)}{2\sqrt{r}}\frac{  (1+r) \beta_{-} f_{n-1}^{(q)} - (1-r) \beta_{+} e_{n-1}^{(q)}}{2(n+q) \beta_{+} \beta_{-} - \frac{1}{2} - E} \nonumber \\
			&& + \frac{\left[ \frac{\Delta}{2} - 2g^{2}(1-r^{2}) (n+q) \right] f_{n}^{(q)}}{2(n+q) \beta_{+} \beta_{-} - \frac{1}{2} - E},	
			\label{recur_tpQRM_en}
		\end{eqnarray}
		\begin{eqnarray}
			f_{n+1}^{(q)} &=& (1+r) \beta_{-} \frac{ (1-r) \beta_{+} e_{n-1}^{(q)} - (1+r) \beta_{-} f_{n-1}^{(q)} }{ 16r \left( n+q+\frac{1}{4} \right)\left( n+q+\frac{3}{4} \right) } \nonumber \\
			&& + \frac{ \left[ 2(n+q) \beta_{-} (2-\beta_{+}^{2}) - \left( \frac{1}{2} + E \right) \beta_{+} \right] f_{n}^{(q)} }{ 8\sqrt{r}g \left( n+q+\frac{1}{4} \right)\left( n+q+\frac{3}{4} \right) } \nonumber \\
			&& + \frac{ \left[-\frac{\Delta}{2} - 2g^{2} (1-r^{2}) (n+q) \right] \beta_{+} e_{n}^{(q)}}{ 8\sqrt{r}g \left( n+q+\frac{1}{4} \right)\left( n+q+\frac{3}{4} \right) }.
			\label{recur_tpQRM_fn}
		\end{eqnarray}
	\end{subequations}
	
	Transforming back to the original Hamiltonian, the eigenfunction is $\vert \Psi^{(q)} \rangle = P^{-1} S(-\theta)  \left\vert \psi^{(q)} \right\rangle$. If $\vert\Psi^{(q)}\rangle$ is an eigenfunction of the conserved parity operator $\Pi$, meaning that the corresponding energy is not degenerate, we have
	\begin{eqnarray}
		\Pi \vert \Psi^{(q)} \rangle &=& P^{-1} \left( -\sigma_{x} \otimes e^{i \frac{\pi}{2} a^{\dagger}a }\right) S(-\theta) \left\vert \psi^{(q)} \right\rangle \nonumber \\
		&\propto& \vert \Psi^{(q)} \rangle.
		\label{pro_relation}
	\end{eqnarray}
	Using
	\begin{eqnarray}
		S(\theta) \vert q,0 \rangle = \frac{1}{\cosh^{2q} \theta} \sum_{n=0}^{+\infty} \frac{\sqrt{\left[ 2 \left(n+q-\frac{1}{4} \right)\right]!}}{2^{n} n!} \tanh^{n} \theta \vert q,n \rangle, \nonumber	
	\end{eqnarray}
	the projection of both sides of Eq.~\eqref{pro_relation} onto the vacuum state $\left\vert q,0 \right\rangle$, yields the $G$-function
	\begin{equation}
		G_{\pm}^{(q)}(E) = \sum_{n=0}^{ +\infty } \left( e_{n}^{(q)} \pm f_{n}^{(q)} \right) \frac{ \left[ 2\left( n+q-\frac{1}{4} \right) \right] !}{2^{n} n!} \tanh^{n} {\theta},
		\label{gfunc-aniso}
	\end{equation}
	where $e_{n}^{(q)}(E)$ and $f_{n}^{(q)}(E)$ are determined by the recurrence relation Eq.~\eqref{recur_tpQRM} with the initial condition $f_{0}^{(q)}(E) = 1$. The label $\pm $ denotes even/odd parity within each Bargmann subspace $\mathcal{H}_q$ where the $\mathbb{Z}_4$-symmetry reduces to $\mathbb{Z}_2$. Zeros of the G-functions will give all eigenenergies of the atpQRM. Our $G$-function provides a more concise form than that in \cite{cui_exact_2017}, though they are basically equivalent. Furthermore, for the isotropic case $r=1$, the recursive relation Eq.~\eqref{recur_tpQRM} simplifies to
	\begin{subequations}
		\begin{eqnarray}
			e_{n}^{(q)} = \frac{\frac{\Delta}{2} f_{n}^{(q)}}{2(n+q) \beta_{+} - \frac{1}{2} - E},
		\end{eqnarray}
		\begin{eqnarray}
			f_{n+1}^{(q)} &=& \frac{-\frac{\Delta}{2} \beta_{+} e_{n}^{(q)} + \left[ 2(n+q) \left( 1+4g^{2} \right) - \left( \frac{1}{2} + E \right) \beta_{+} \right] f_{n}^{(q)}}{8g \left( n+q+\frac{1}{4} \right)\left( n+q+\frac{3}{4} \right)} \nonumber \\
			&& -\frac{f_{n-1}^{(q)}}{4 \left( n+q+\frac{1}{4} \right)\left( n+q+\frac{3}{4} \right)},
		\end{eqnarray}
	\end{subequations}
	with $\beta_{+} = \sqrt{1 - 4g^{2}}, \beta_{-} = 1$. Then the $G$-function is identical to that for isotropic tpQRM which was found heuristically in \cite{chen_exact_2012} using the present approach. It was rigorously proven in \cite{braak_spectral_2023} that the zeros of this $G$-function indeed give all normalizable solutions of the Schr\"odinger equation and thus the pure point spectrum of $H$ in the isotropic case. Because the essential structure of the Hamiltonian in terms of its singular points in the Bargmann representation is the same for the isotropic and the present anisotropic model, we may infer that the $G$-function \eqref{gfunc-aniso} yields the complete pure point spectrum of \eqref{H_orgin}.
	
\begin{figure}[tbp]
		\includegraphics[width=\linewidth]{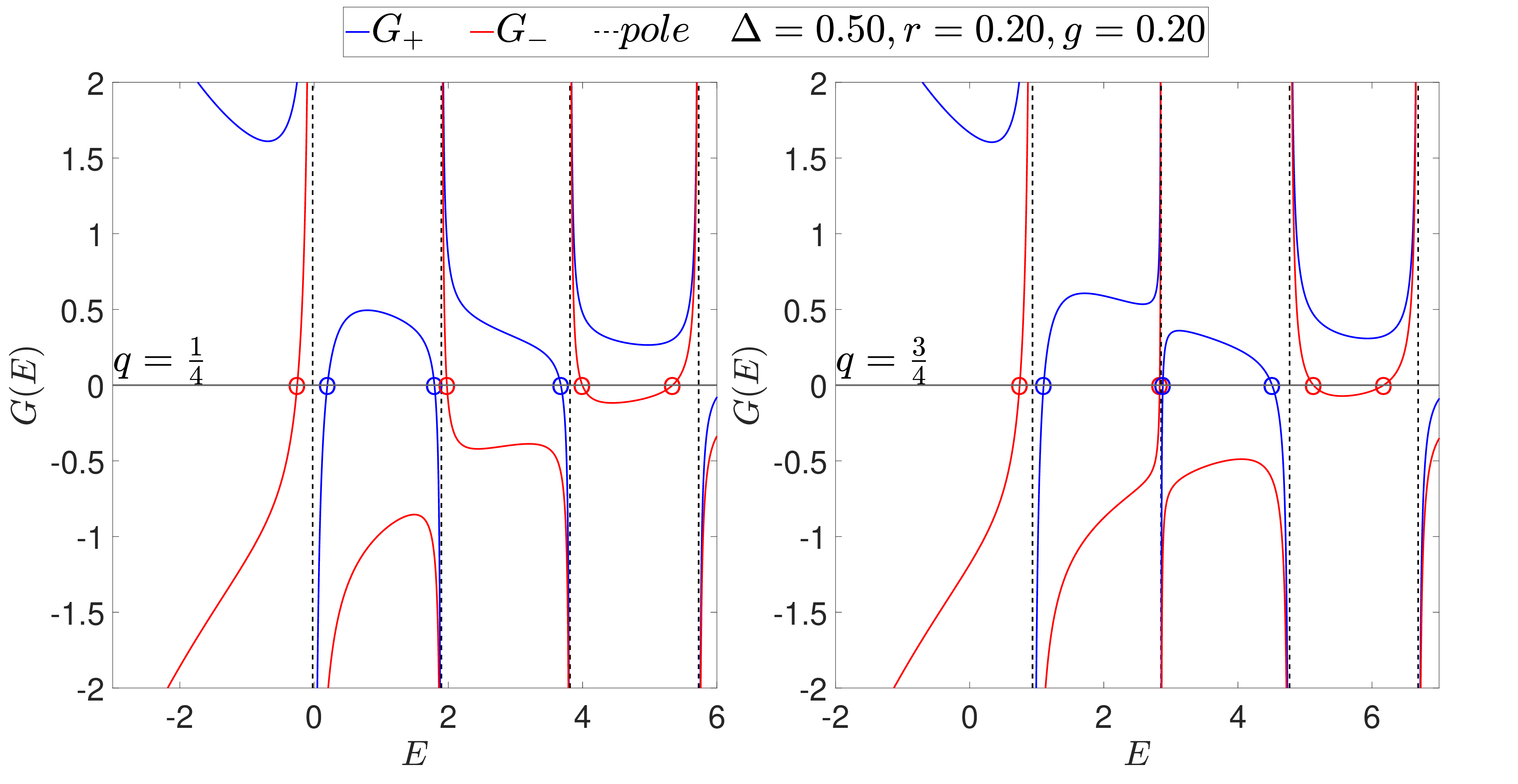}
		\caption{G-curves at $\Delta =0.50$, $r=0.20$, and $g=0.20$ as functions of $E$ for $q=1/4$ (left panel) and $q=3/4$ (right panel).The blue (red) lines indicate $G_{+}^{(q)}$ ($G_{-}^{(q)}$) curves, the black dashed lines denote $E_{n}^{(q,pole)}$, and the opened circles represent the zeros.}
		\label{G_tpQRM_fig}
	\end{figure}

	Note that the recursion for $e_{n}^{(q)}$ in Eq.~\eqref{recur_tpQRM_en} yields the pole structure
	\begin{eqnarray}
		E_{n}^{(pole)} = 2(n+q) \beta_{+} \beta_{-} - \frac{1}{2},
		\label{pole}
	\end{eqnarray}
	where $G_{\pm }^{(q)}$ diverges, as shown in Fig.~\ref{G_tpQRM_fig}.
	
\begin{figure}[tbp]
		\includegraphics[width=\linewidth]{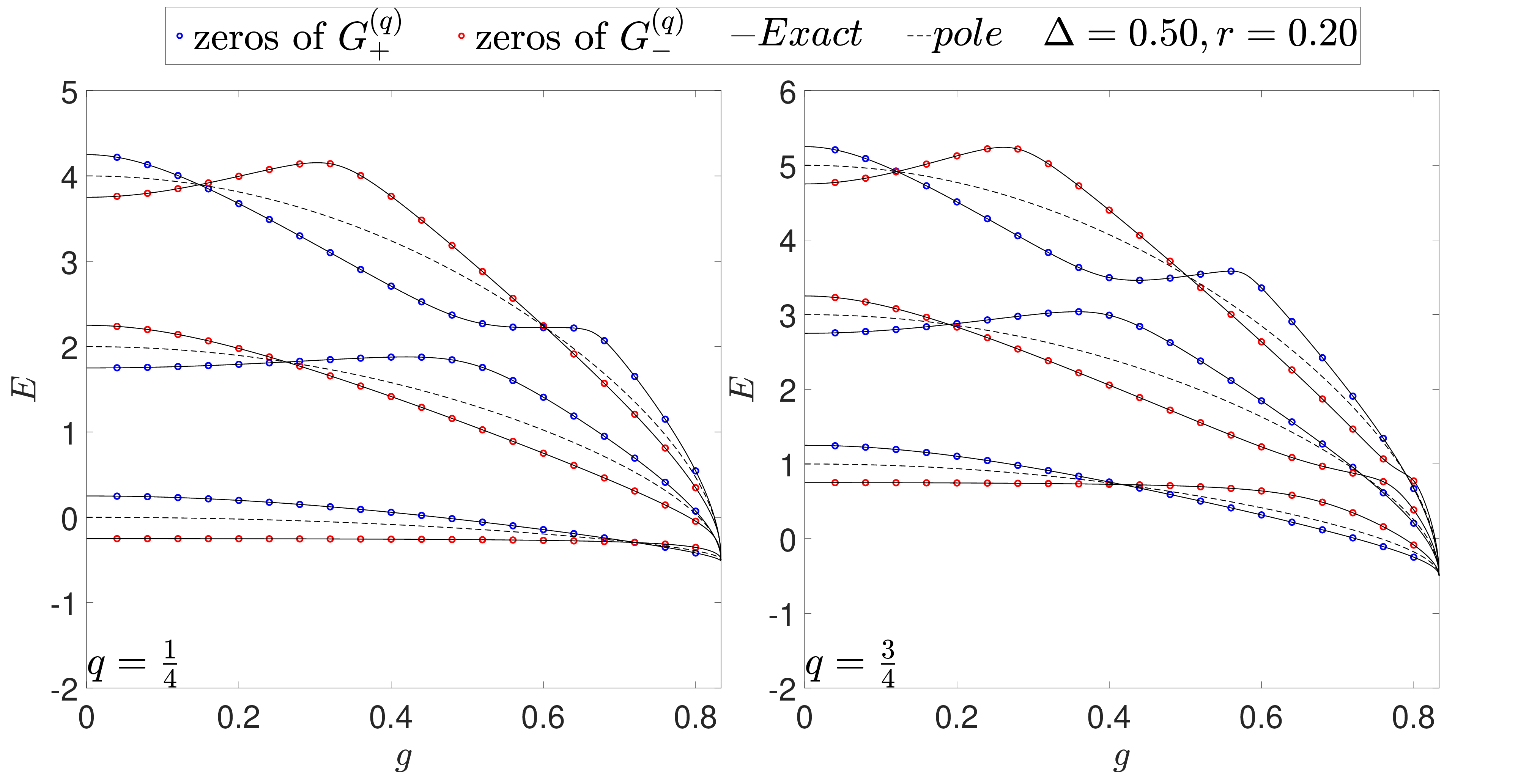}
		\caption{Energy spectra at $\Delta =0.50$ and $r=0.20$ as functions of $g$ for $q=1/4$ (left panel) and $q=3/4$ (right panel). The black solid lines represent eigenvalues obtained by exact diagonalization (ED), and the dashed lines denote $E_{n}^{(q, pole)}$. Blue circles mark zeros of $G_{+}^{(q)}$, while red circles mark zeros of $G_{-}^{(q)}$.}
		\label{spectra}
	\end{figure}

	As shown in Fig.~\ref{spectra}, the zeros of $G$-function accurately reproduce the energy spectra obtained by numerical exact diagonalization if $g$ stays below the critical point $g_c$. Two adjacent levels with different parities cross at the pole lines, indicating doubly degenerate eigenvalues. The energy spectra seem to ``collapse" to a single energy at $g_{c}$. Precisely at this collapse coupling, $\beta_{+} = 0$, rendering the Bogoliubov transformation singular ($\theta=\infty$). We will later demonstrate the asymptotic behavior as $g \rightarrow g_{c}$.
	
	\section{Doubly degenerate eigenvalues}
	
	\begin{figure}[tbp]
		\includegraphics[width=0.9\linewidth]{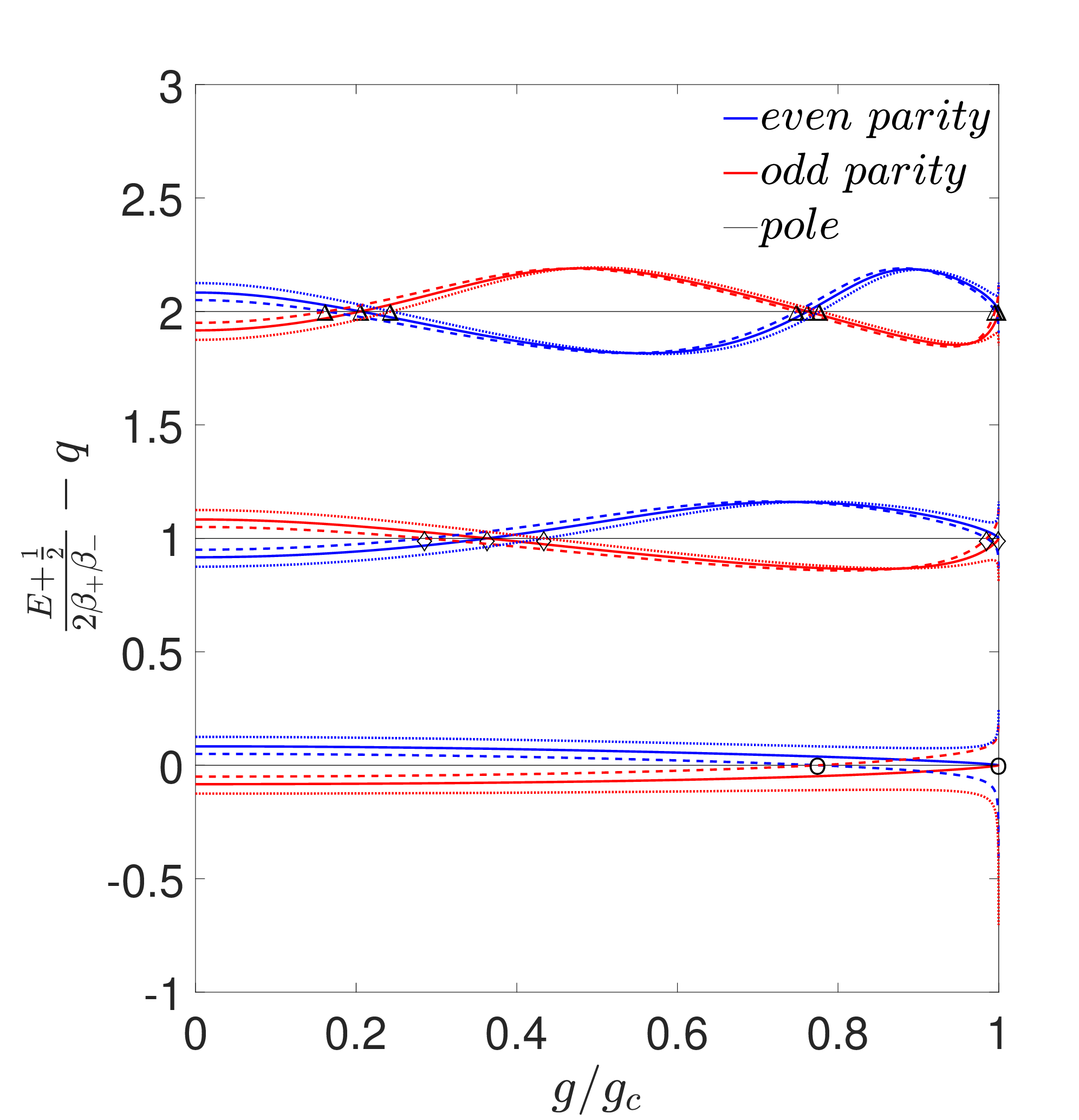}
		\caption{The scaled energy spectra $E^{\prime} = (E+1/2)/(2\beta_{+}\beta_{-}) - q$ as a function of $g$ (normalized by $g_{c}$) for $r=0.25$, $q=1/4$. Blue (red) lines mark even (odd) parity levels, and black lines denote pole lines. Points marked by circles, diamonds, and triangles denote the zeros of $F_{0}^{(q)}(g)$, $F_{1}^{(q)}(g)$ and $F_{2}^{(q)}(g)$, respectively. Solid lines correspond to $\Delta =0.60$, dashed lines to $\Delta =0.50$, and dotted lines to $\Delta =0.70$. The truncation number for Fock space in the exact diagonalization is $5000$.}
		\label{spectra_fng_fig}
	\end{figure}
	
	According to the pole structure in Eq.~\eqref{pole}, we immediately know that the spectral collapse also occurs in the atpQRM, similar to its isotropic counterpart. The first pole is located at $E_{0}^{(pole)} = 2q \beta_{+} \beta_{-} - 1/2$. The distance between adjacent poles is $2\beta_{+} \beta _{-}$, which vanishes as $\beta_{+} \rightarrow 0$, i.e., $g \rightarrow g_{c}$. All zeros of $G_{\pm}^{(q)}$ (energy levels) are constrained between adjacent poles with intervals of length $2\beta_{+} \beta_{-}$; therefore their distance goes to zero as $g\rightarrow g_c$. The poles fill the whole real axis above the threshold energy $E_{c}=-1/2$ densely as $g \rightarrow g_{c}$; thus a \emph{continuum} of states emerges at $g=g_c$.
	
	However, as shown in Fig.~\ref{spectra}, some discrete energy levels remain below $E_{0}^{(pole)}$ as $g\rightarrow g_c$, and thus are not pinched between two poles. These states form the discrete set of bound states below the continuum threshold $E_c$ at the collapse point.
	
	When $E = E_{n}^{(pole)}$ belongs to the spectrum, it corresponds to the exceptional part, not given by a zero of the $G$-function. In this case, the zero in the denominator of the recurrence relation for $e_n^{(q)}$ is lifted by a corresponding zero in the numerator. From \eqref{recur_tpQRM_en} we obtain two cases,
	\begin{eqnarray}
		\frac{g (1-r^{2})}{2\sqrt{r}} \beta_{-} f_{n-1}^{(q)} - \frac{g (1-r)^{2}}{2 \sqrt{r}} \beta_{+} e_{n-1}^{(q)} \nonumber \\
		+ \left[ \frac{\Delta}{2} - 2g^{2}(1-r^{2}) (n+q) \right] f_{n}^{(q)} = 0,
		\label{num_en}
	\end{eqnarray}
	or
	\begin{eqnarray}
		f_{n}^{(q)} = f_{n-1}^{(q)} = e_{n-1}^{(q)} = 0, \nonumber
	\end{eqnarray}
	which correspond to a doubly degenerate eigenvalue and a special nondegenerate state respectively. Both states belong to the exceptional spectrum of the atpQRM. For a doubly degenerate eigenvalue, we can define a function $F_{n}^{(q)}(g)$ by eliminating $e_{n-1}^{(q)}$ in Eq.~\eqref{num_en},
	\begin{eqnarray}
		F_{n}^{(q)}(g) &=& \sum_{i=0}^{n} \frac{f_{i}^{(q)}}{(n-i)!} \left[ \frac{ g (1-r)^{2} }{ 4 \sqrt{r} \beta_{-} }\right]^{n-i} \times  \nonumber \\
		&& \left[ \frac{\Delta}{2} + \frac{1+r}{1-r} 2(n-i) - 2g^{2} \left( 1-r^{2} \right) (n+q) \right] . \nonumber \\
		\label{Fng}
	\end{eqnarray}
	The coupling strength $g_{n}^{(q)}$, where the doubly degenerate state on the $n$th pole line occurs, can be determined by solving $F_{n}^{(q)}(g) = 0$ for given $\Delta$ and $r$, as illustrated in Fig.~\ref{spectra_fng_fig} where the exact diagonalization obtains the curves.
	
	First, we assume that only one state is located below $E_0^{pole}$ which is thus the ground state. The first excited state lies then above the first pole line. Therefore, the crossing point of the ground state and the first excited state in the energy spectra is located at
	\begin{eqnarray}
		g_{0}^{(q)} &=& \frac{1}{2}        \sqrt{\frac{\Delta}{ q (1-r^{2})}}, \nonumber \\
		E_{0}^{(q)} &=& 2q \sqrt{ \left(1 - \frac{\Delta}{4q} \frac{1-r}{1+r} \right) \left( 1 - \frac{\Delta}{4q} \frac{1+r}{1-r} \right)} - \frac{1}{2}.
	\end{eqnarray}
	For $g_{0}^{(q)} \leq 1/(1+r)$ and $E_{0}^{(q)}$ to be real, it is required that $\Delta \leq 4q \frac{1-r}{1+r}$. There is thus a critical $\Delta_c^{(q)}$
	\begin{eqnarray}
		\Delta_{c}^{(q)} = 4q \frac{1-r}{1+r}, \nonumber
	\end{eqnarray}
	above which the crossing does not occur in the region $g<g_c$.
	When $ \Delta = \Delta_{c}^{(q)}$, the crossing point occurs at $g_{0}^{(q)} = g_{c}$ and $E_{0}^{(q)}=E_{c}$. This implies that for $\Delta = \Delta_{c}^{(q)}$, the ground state and the first excited state are expected to intersect at the collapse point $g_{c}=1/(1+r)$, $E_{c}=-1/2$. In the isotropic case, $F_{n}^{(q)}(g)$ simplifies to $F_{n}^{(q)}(g) = \Delta f_{n}^{(q)} / 2$. Because $f_{0}^{(q)} = 1$, the ground state and the first excited state intersect only when $\Delta=0$, i.e. $\Delta_{c}^{(q)}$ for $r=1$. There is no first-order phase transition in the isotropic tpQRM for any $\Delta>0$, similar to the linear QRM \cite{chen_multiple_2021}.
	
	For $n>0$ the pole lines are crossed more than once. From the solid line for $r=0.25$, $\Delta = \Delta_{c}^{(1/4)} = 0.60$ in Fig.~\ref{spectra_fng_fig}, we observe numerically that the crossing points closest to $g_c$ seem to be located at $g_c$ for $\Delta_c^{(1/4)}$.
	Interestingly, this can be obtained analytically by continuing the recurrence relations to $g_c$ although they are not defined directly at $g=g_{c}$. The recursive relation Eq.~\eqref{recur_tpQRM_fn} simplifies to
	\begin{eqnarray}
		f_{n+1}^{(q)} = \frac{ (n+q) f_{n}^{(q)} - \frac{1}{4} f_{n-1}^{(q)}}{ \left( n+q+\frac{1}{4} \right)\left( n+q+\frac{3}{4} \right) }, \nonumber
	\end{eqnarray}
	which is independent of $E$, $r$ and $\Delta$. Starting from $f_{0}^{(q)} = 1$, we obtain
	\begin{eqnarray}
		f_{n}^{c} &=& (2^{n} n!)^{-1},
		\label{fnc}
	\end{eqnarray}
	independent of $E$, $q$, $r$ and $\Delta$. Substituting $\{f_{n}^{c}\}$ and $\Delta_{c}^{(q)}$ into $F_{n}^{(q)}(g)$ Eq.~\eqref{Fng} yields,
	\begin{eqnarray}
		F_{n} \left( g_{c} \right) &=& - \sum_{i=0}^{n} \frac{2i}{2^{i} i! (n-i)!} \left[ \frac{(1-r)^{2}}{8r} \right]^{n-i} \frac{1+r}{1-r} \nonumber \\
		&& + \sum_{i=0}^{n} \frac{n}{2^{i} i! (n-i)!} \left[ \frac{(1-r)^{2}}{8r} \right]^{n-i} \frac{8r}{1-r^{2}}\nonumber \\
		&=& - \frac{1+r}{1-r} \frac{1}{(n-1)!} \left[ \frac{(1+r)^{2}}{8r} \right]^{n-1} \nonumber \\
		&& + \frac{8nr}{1-r^{2}} \frac{1}{n!} \left[ \frac{(1+r)^{2}}{8r} \right]^{n} = 0.
	\end{eqnarray}
	Because the doubly degenerate solution occurs for any $n$, we infer that all energies belonging to the spectral continuum above $E_c$ at $g_c$ are doubly degenerate for fixed $q$. This will be confirmed by the analysis of the effective Schr\"odinger equation at the collapse point given in section~\ref{sec-4}.
	
\begin{figure}[tbp]
		\includegraphics[width=\linewidth]{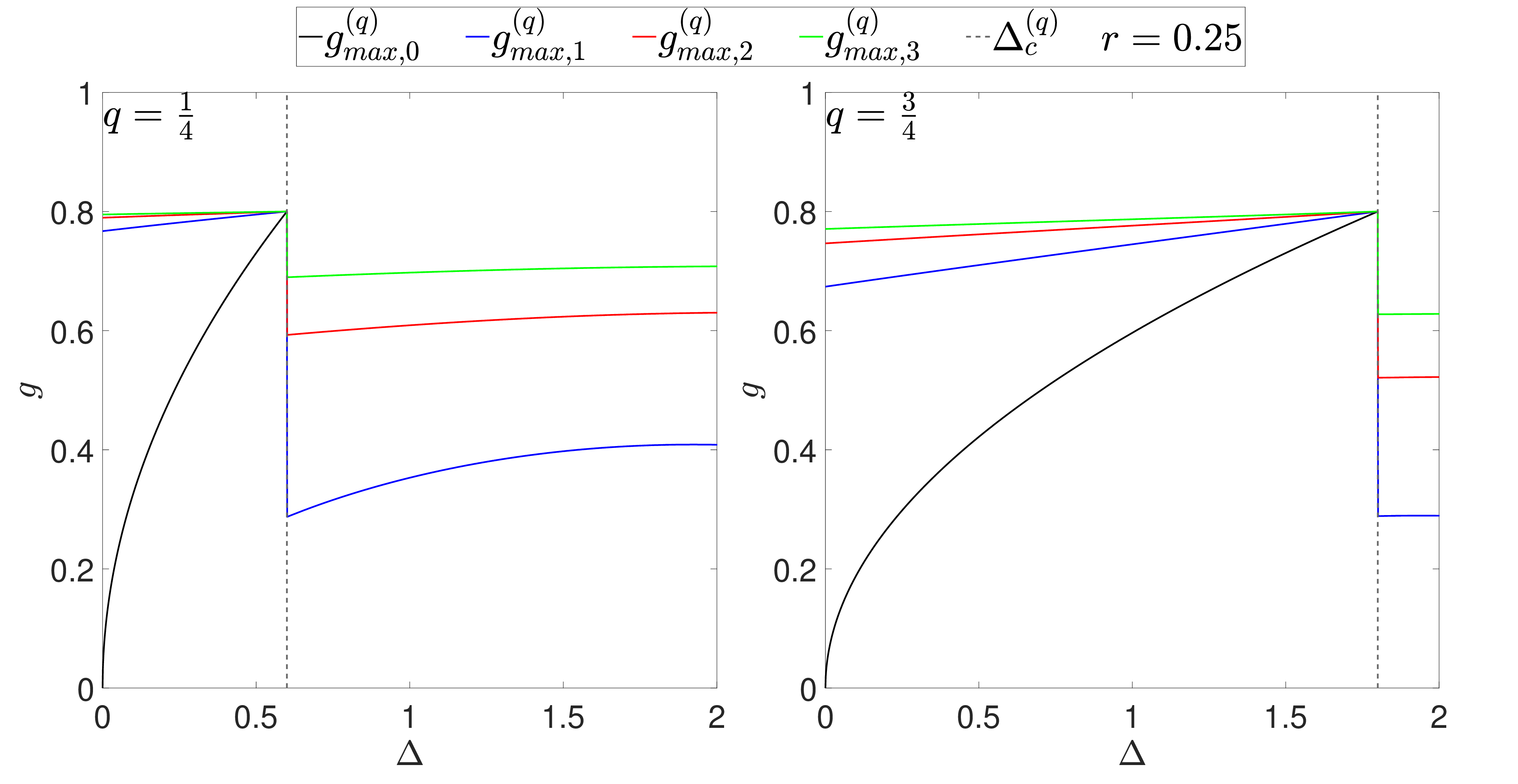}
		\caption{The last crossing point  $g_{max,n}^{(q)}$ in the $g/\Delta $ plane at $r=0.25$ for $q=1/4$ (left panel), and $q=3/4$ (right panel) is given by the maximum among the solutions $g_{n}^{(q)}$ of $F_{n}^{(q)}(g)=0$ for fixed $n$. In this diagram, the black, blue, red, and green lines correspond to $F_{0}^{(q)}(g)$, $F_{1}^{(q)}(g)$, $F_{2}^{(q)}(g)$ and $F_{3}^{(q)}(g)$, respectively. The dashed line indicates $\Delta_{c}^{(q)}$ and $g_c=0.8$. For $\Delta>\Delta_c^{(q)}$, no doubly degenerate exceptional point is located on the pole line $m=0$.}
		\label{phase_gm_fig}
	\end{figure}

	If $\Delta$ deviates slightly from $\Delta_{c}^{(q)}$ by a very small value $\delta$, we have to first order in $\delta$,
	\begin{eqnarray}
		F_{n}\left( g_{c} \right) =
		\frac{\delta}{2} \left[ \frac{(1+r)^{2}}{8r} \right] ^{n}. \nonumber
	\end{eqnarray}
	Thus, the last crossing point will be shifted from $g_{c}$. Only when $\Delta = \Delta_{c}^{(q)}$ does $F_{n}$ equal zero at $g_{c}$. Fig.~\ref{phase_gm_fig} shows that the last crossing point $g_{max,n}^{(q)}$ from $F_{n}^{(q)}(g)=0 $ in Eq.~\eqref{Fng} for any $n$ is discontinuous at $\Delta_{c}^{(q)}$ ($0.60$ in left panel for $q=1/4$ and $1.80$ in right panel for $q=3/4$) and approaches $g_{c} = 0.80$.
	
	If $\Delta$ is not exactly at $\Delta_{c}^{(q)}$, the last crossing points will not be at the collapse point, as demonstrated by the dashed and dotted lines in Fig.~\ref{spectra_fng_fig} for $q=1/4$. Additionally, if $\Delta < \Delta_{c}^{(q)}$, in the interval between the last crossing point $g_{max,n}^{(q)}$ and the collapse point $g_{c}$, the energy levels with odd parity lie above the pole lines, while those with even parity lie below. The situation reverses for $\Delta >\Delta _{c}^{(q)}$. This phenomenon suggests a first-order phase transition between the ground state and the first excited state when $\Delta < \Delta_{c}^{(q)}$. The downward/upward bending of the curves in Fig.~\ref{spectra_fng_fig} for $g$ very close to $g_c$ is a numerical artifact due to the finite-dimensional state space used in exact diagonalization.
	
	The previously mentioned special non-degenerate points can be used to estimate the number of bound states at the collapse point as follows. First, we derive the exceptional $G$-function whose zeros give the non-degenerate solutions on the $m^{th}$ pole line, i.e. with energy $E = 2(m+q) \beta_{+} \beta_{-} - 1/2 $. We have then
	\begin{equation*}
		f_{m}^{(q)} = f_{m-1}^{(q)} = e_{m-1}^{(q)} = 0.
	\end{equation*}
	From the recursive relation Eq.~\eqref{recur_tpQRM}, one immediately finds that all $f_{n \leq m}$ and $e_{n<m}$ are zero. Now, both the numerator and dominator in Eq.~\eqref{recur_tpQRM_en} for $e_{m}^{(q)}$ are zero, so the value of $e_{m}^{(q)}$ can be arbitrary. By setting $e_{m}^{(q)}=1$, the recursive relation remains unchanged, the difference is that now recursion starts from $e_{m}^{(q)}$ and $E$ is set to be $E = 2(m+q) \beta_{+} \beta_{-} - 1/2$.  As a result, the special $G$-function for special non-degenerate points is rewritten as a function of $g$:
	\begin{eqnarray}
		G_{\pm}^{(q)}(m,g) && = \frac{ \left[ 2\left( m+q-\frac{1}{4} \right) \right] !}{2^{m} m!} \tanh^{m} {\theta} + \nonumber \\
		&& \sum_{n=m+1}^{ +\infty } \left( e_{n}^{(q)}(g) \pm f_{n}^{(q)}(g) \right) \frac{ \left[ 2\left( n+q-\frac{1}{4} \right) \right] !}{2^{n} n!} \tanh^{n} {\theta},
		\nonumber \\
		\label{exc-gfunc}
	\end{eqnarray}
	whose zeros give the special non-degenerate points. Because $E$ is fixed to $2(m+q) \beta_{+} \beta_{-} - 1/2$, the dominator in Eq.~\eqref{recur_tpQRM_en} for all $e_{n>m}^{(q)}$ is $2(n-m)\beta_{+} \beta_{-} \neq 0$ except for $g_{c}$. So $G_{\pm}^{(q)}(g)$ is well defined when $g<g_{c}$.
	
\begin{figure}[tbp]
		\includegraphics[width=\linewidth]{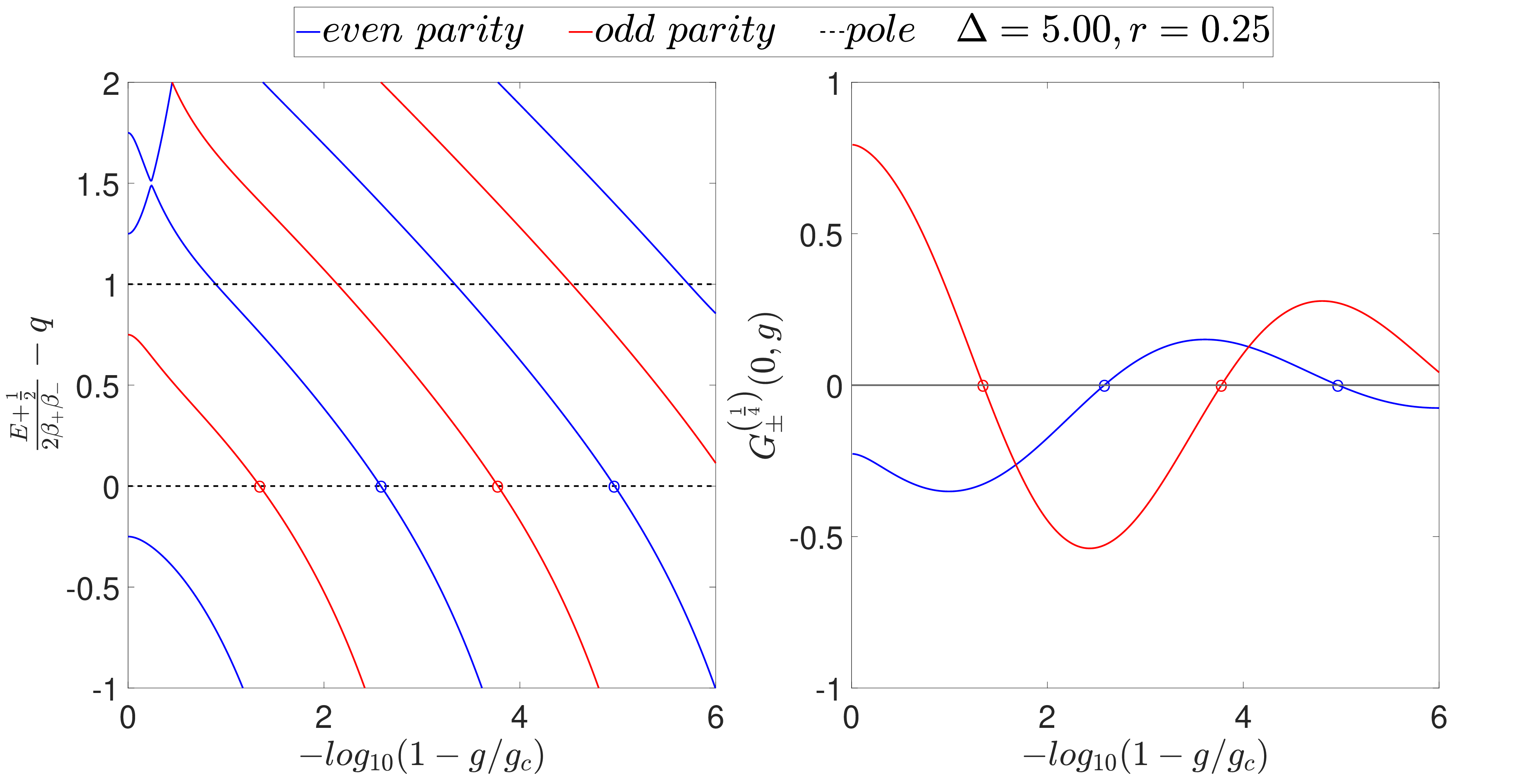}
		\caption{The left panel presents the scaled spectra $E^{\prime} = (E+1/2)/ (2\beta_{+}\beta_{-}) - q$ while the right panel shows $G_{\pm}^{(1/4)}(0,g)$, both as functions of $g$ with parameters $\Delta=5.00$, and $r=0.25$. The $x$-axis is scaled logarithmically, $x=-\log_{10} (1-g/g_{c})$. Blue (red) lines indicate even (odd) parity, and dashed lines indicate pole lines. The special non-degenerate points have been circled with corresponding colors.}
		\label{snd_m0_fig}
	\end{figure}

	For any $g<g_c$, the level lines in the spectral graph as function of $g$ below the $0^{th}$ pole line are either continuously connected to the point $g=0$ or they lie above the $0^{th}$ line for all $g<g_i<g_c$, cross this line at $g_i$ and stay below the $0^{th}$ pole line for all $g>g_i$. Because this line has the value $E_c$ at $g_c$ (see the left panel of Fig.~\ref{snd_m0_fig}), all these level lines end in discrete bound states below $E_c$ at $g_c$. The number of those bound states can therefore be obtained by counting the zeros of  $G_{\pm}^{(q)}(0,g)$ for $g<g_c$. If one adds the bound states computed for $g=0$ (always below the $0^{th}$ pole line), one may deduce the number of bound states (but not the exact energies) right at the collapse point, although the $G$-function from \eqref{gfunc-aniso} is not defined there.
	
\begin{figure}[tbp]
		\includegraphics[width=1.0\linewidth]{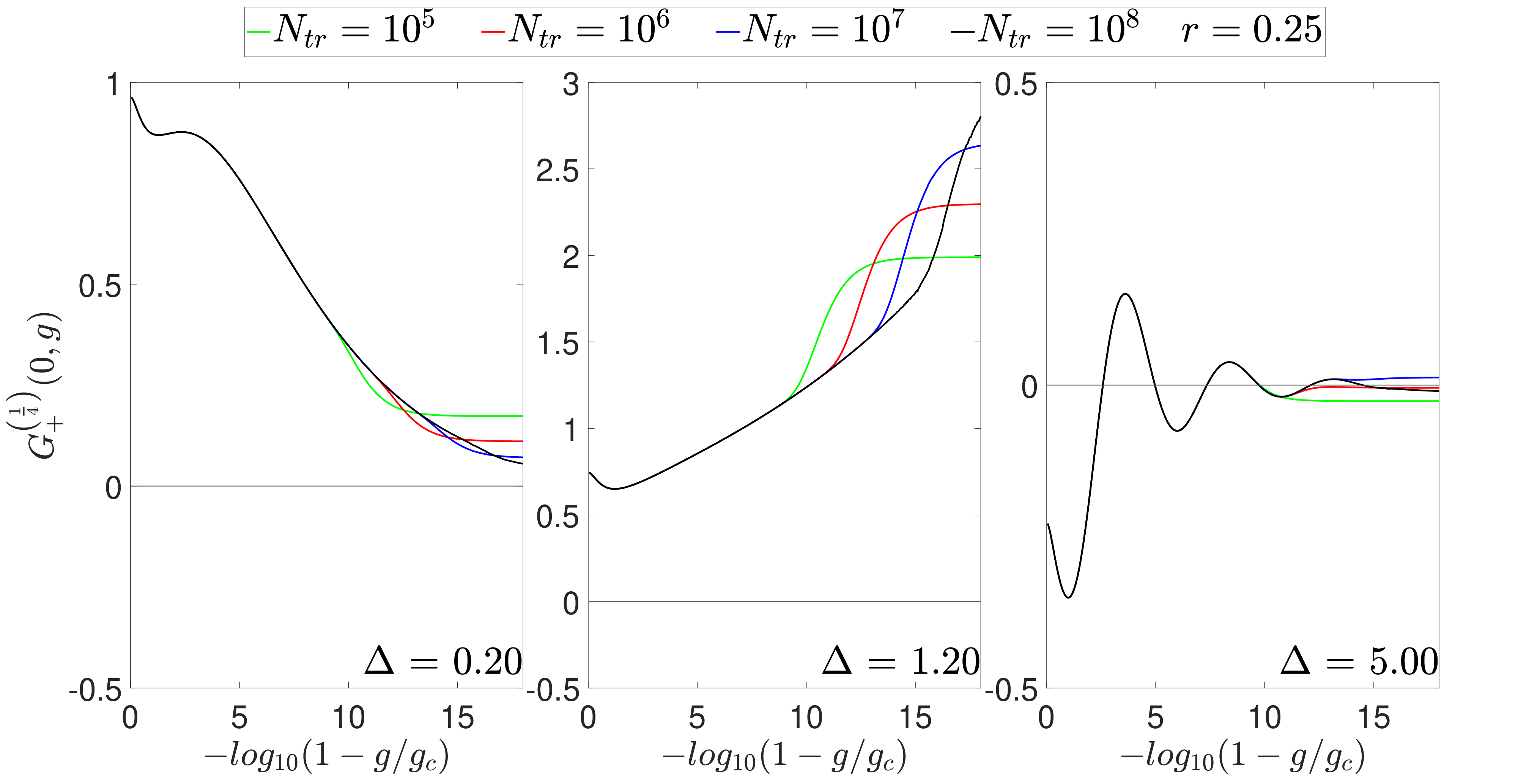}
		\caption{$G_+^{(1/4)}(0,g)$ at $r=0.25$ and $\Delta=0.20$ (left), $\Delta =1.20$ (middle) and $\Delta=5.00$ (right). The $x$-axis is scaled logarithmically, $x=-\log_{10} (1-g/g_{c})$. The green, red, blue, and black lines indicate that the series in \eqref{exc-gfunc} is truncated at $N_{tr} = 10^{5},10^{6},10^{7},10^{8}$, respectively.}
		\label{G_nond_fig}
	\end{figure}

	Figure.~\ref{G_nond_fig} displays the exceptional $G$-function for $\Delta<\Delta^{(1/4)}_c$ (left), $\Delta_c^{(1/4)}<\Delta<\Delta_c^{(3/4)}$ (middle) and $\Delta>\Delta_c^{(3/4)}$ (right). A logarithmic scale for $g$ is used to resolve the exponentially close vicinity of $g_c$. One sees that the numerics of the $G$-function is stable but, as expected, will yield the correct result close to $g_c$ only for very large truncation number of the series expansion. Especially, it is not possible to determine numerically whether $G_+^{(1/4)}(0,g)$ has a zero for $\Delta<\Delta_c^{(1/4)}$, whereas it is clear that no zero exist for  $\Delta_c^{(1/4)}<\Delta<\Delta_c^{(3/4)}$. The numerics also indicate a probably infinite number of zeroes which are clustered exponentially close to the collapse point for large $\Delta$. These findings will be confirmed by analytical arguments in the next sections.  We note that the numerical acquisition of the spectrum via exact diagonalization is numerically much more demanding than the evaluation of the special $G$-function close to $g_c$. Therefore Fig.~\ref{snd_m0_fig} only presents results up to $1-g/g_{c} \sim 10^{-6}$.
	
	In the isotropic case ($r=1$) we have $\Delta_{c}^{(q)}=0$. In this (trivial) case, all levels lie above $E=-1/2$ and a spectral continuum emerges at $g_c=1/2$ with no bound states. We shall show now that also for the anisotropic case, there is no discrete spectrum at $\Delta_c^{(1/4)}$, whereas it exists for $\Delta\neq\Delta_c^{(1/4)}$.
	
	\section{Absence of the bound spectrum for $\Delta_c^{(1/4)}$}\label{sec-4}
	
	We show first that for all values of the anisotropy $r$, the threshold energy $E_c=-1/2$ is a doubly degenerate element of the continuous spectrum for parameters $\Delta=\Delta_c^{(1/4)}$ and $g=g_c$.
	After a rotation in pseudo-spin space, the Hamiltonian Eq.~\eqref{H_orgin} can be written
	\begin{equation}
		H_r=
		\left(\!
		\begin{array}{cc}
			\hat{x}^2 & \frac{1-r}{1+r}i\hat{x}\hat{p}\\
			-\frac{1-r}{1+r}i\hat{p}\hat{x} & \hat{p}^2
		\end{array}
		\!\right)
		-\frac{1}{2}\id,
		\label{ham-crit}
	\end{equation}
	where $\hat{x}=(a+a^\dagger)/\sqrt{2}$ and $\hat{p}=i(a^\dagger-a)/\sqrt{2}$ are the usual self-adjoint position and momentum operators. The ``generalized" vectors
	\begin{equation}
		|\Psi_\alpha\rangle=
		\left(
		\begin{array}{c}
			\psi_x\\
			\alpha\psi_p
		\end{array}
		\right)
		\label{0-states}
	\end{equation}
	yield formally $H_r|\Psi_\alpha\rangle=E_c|\Psi_\alpha\rangle$, if
	$\hat{x}\psi_x=\hat{p}\psi_p=0$. The value of $\alpha$ is arbitrary, therefore $E_c$ is a doubly degenerate element of the continuous spectrum of $H_r$: The spectra of $\hat{x}$ and $\hat{p}$ are purely continuous, and $\psi_x,\psi_p$ are not normalizable.
	
	Moreover, the operator $H_0=H_r+(1/2)\id$ is positive semi-definite. For any element of $L^2(\mathbb{R})\otimes \mathbb{C}^2$, $|\Psi\rangle=(\psi_1,\psi_2)^T$, we have
	\begin{align}
		\langle\Psi|H_0|\Psi\rangle&=
		\langle\psi_1|\hat{x}^2|\psi_1\rangle
		+\Delta_c^{(1/4)}i\langle \psi_1|\hat{x}\hat{p}|\psi_2\rangle\nn\\
		&-\Delta_c^{(1/4)}i\langle\psi_2|\hat{p}\hat{x}|\psi_1\rangle
		+\langle\psi_2|\hat{p}^2|\psi_2\rangle\nn\\
		&= \langle\Phi|\Phi\rangle +\left(1-(\Delta_c^{1/4})^2\right)
		\langle\phi|\phi\rangle \ge 0,
		\label{pos-def}
	\end{align}
	with $|\Phi\rangle=\hat{x}|\psi_1\rangle+i\Delta_c^{(1/4)}\hat{p}|\psi_2\rangle$ and $|\phi\rangle=\hat{p}|\psi_2\rangle$. The last line in \eqref{pos-def} follows because $\Delta_c^{(1/4)}\le 1$. Indeed, the only case where $H_0$ is not strictly positive definite corresponds to the RWA-limit (rotating wave approximation) $r=0$, as we show next.
	
	At $r=0$, the count-rotating terms vanish, allowing the Hamiltonian to be directly diagonalized in the basis $\vert \uparrow, 2n \rangle$ and $\vert \downarrow, 2n+2 \rangle$,
	\begin{eqnarray}
		E_{n} = 2n+1 \pm \sqrt{ \left( 1 - \frac{\Delta}{2} \right)^{2} + g^{2} (2n+1) (2n+2)}. \nonumber
	\end{eqnarray}
	When $g_{c}=1$ with $\Delta_{c}^{(1/4)}=1$ or $\Delta_{c}^{(3/4)}=3$, the eigenvalues are $E_{n} = -\frac{1}{2}$ and $4n + 5/2$. In this case a spectral collapse in the intuitive sense indeed happens at $g=g_c$ and for $\Delta$ either $\Delta_c^{(1/4)}$ or $\Delta_c^{(3/4)}$, as half of the energy levels lie at $E_{c}$, while the other half forms a discrete spectrum $\{E_n\}$ with $E_n > E_c$. The infinite degeneracy of the energy $E_c$ allows then also to construct the generalized eigenstates \eqref{0-states}. But because there are normalizable states with energy $E_c$ as well, the operator $H_0$ is not strictly positive definite for $r=0$. For $\Delta<1$, half of the energy levels are below $E_{c}$, while the others are above $E_{c}$. When $1<\Delta<3$, all energy levels are above $E_{c}$. For $\Delta>3$, again, half of the energy levels are below $E_{c}$, and the others are above $E_{c}$.
	
	The case $r=1$ is likewise simple at $\Delta_c^{(1/4)}$ and $g_c$. From
	\eqref{ham-crit}, we see that the Hamiltonian $H_0$ equals $\hat{x}^2$ for spin-up and $\hat{p}^2$ for spin-down. The continuous spectrum stretches therefore from $E_c=-1/2$ to positive infinity and is doubly degenerate.
	
	For all other values of the anisotropy $0<r<1$, the Hamiltonian $H_0$ is positive definite. Because a continuum emerges at the collapse point $g_c$ for any $r$ above $E_c$, there are no bound states for $\Delta =\Delta_c^{(1/4)}$.
	
	\section{The bound spectrum for $\Delta \neq \Delta_c^{(1/4)}$}\label{sec-4-1}
	
	We consider the Hamiltonian \eqref{H_orgin} at $g=g_c$ for arbitrary $\Delta$ in the $\hat{x}/\hat{p}$-representation as above but with an additional scale factor $\kappa$,
	\begin{equation*}
		\kappa\hat{x}= \frac{a+a^{\dagger}}{\sqrt{2}}, \quad  \kappa^{-1}\hat{p} = i \frac{a^{\dagger}-a}{\sqrt{2}}.
	\end{equation*}
	Using the position basis in $L^2(\mathbb{R})$ in which $\hat{x}$ is multiplication by $\kappa$, the Hamiltonian reads (after rotation in pseudo-spin space)
	\begin{eqnarray}
		H_{c} = \begin{bmatrix}
			\kappa^{2} x^{2} - \frac{1}{2} & - \frac{\Delta}{2} + \frac{1-r}{1+r} \left( x \frac{ \partial }{ \partial x } + \frac{1}{2} \right) \\
			- \frac{\Delta}{2} - \frac{1-r}{1+r} \left( x \frac{ \partial }{ \partial x } + \frac{1}{2} \right) & - \frac{ \partial^{2} }{ \kappa^{2} \partial x^{2}} - \frac{1}{2}
		\end{bmatrix}. \nonumber \\
		\label{coll-H}
	\end{eqnarray}
	The parity operator $\Pi_{p} = -\sigma_{x} \exp\left[ i \frac{\pi}{2} a^{\dagger} a \right]$ exchanges the upper and lower spin component and acts in $L^2(\mathbb{R})$ as Fourier transform via $\exp\left[ i \frac{\pi}{2} a^{\dagger} a \right]$ which maps $\hat{x}$ to $\hat{p}$ and $\hat{p}$ to $-\hat{x}$. The Schr\"{o}dinger equation $ H_{c} \vert \Psi(E) \rangle = E \vert \Psi(E) \rangle $ reads with the definition of  $|\Psi\rangle$ as above,
	\begin{eqnarray}
		\kappa^{2} x^{2} \psi_{1} + \left[ - \frac{\Delta}{2} + \frac{1-r}{1+r} \left( x \frac{ \partial }{ \partial x } + \frac{1}{2} \right) \right] \psi_{2} = \left( E+\frac{1}{2} \right) \psi_{1} \nonumber \\
		\left[- \frac{\Delta}{2} - \frac{1-r}{1+r} \left( x \frac{ \partial }{ \partial x } + \frac{1}{2} \right) \right] \psi_{1}  - \frac{ \partial^{2} }{ \kappa^{2} \partial x^{2} } \psi_{2} = \left( E+\frac{1}{2} \right) \psi_{2}. \nonumber \\
		\label{coll-Schr}
	\end{eqnarray}
	As in the isotropic case, $\psi_1(x)$ is (proportional to) the Fourier transform of $\psi_2(x)$, if the eigenstate $|\Psi(E)\rangle$ has fixed parity.
	Eliminating $\psi_{1}(x)$ yields the following equation for $\psi_2(x)$,
	\begin{eqnarray}
		\left( E + \frac{1}{2} \right) \psi_{2} = - \frac{ \partial }{ \partial x } \left[ \frac{1}{\kappa^{2}} \frac{ \kappa^{2} x^{2} \frac{4r}{(1+r)^{2}} - \frac{1}{2} - E }{ \kappa^{2} x^{2} - \frac{1}{2} - E } \right] \frac{ \partial }{ \partial x } \psi_{2} \nonumber \\
		+ \frac{\Delta^{2} - \left( \frac{1-r}{ 1+r } \right)^{2} }{ 4 \left(\kappa^{2} x^{2} - \frac{1}{2} - E \right)^{2} } \left[ \frac{ \frac{3(1-r)}{1+r} - \Delta }{ \Delta + \frac{1-r}{1+r} } \kappa^{2} x^{2} + \frac{1}{2}  + E \right] \psi_{2}. \nonumber \\
		\label{origin_psi2}
	\end{eqnarray}
	This equation contains the eigenvalue $E$ of the original system \eqref{coll-Schr} also within the differential operators. We can eliminate this dependence by choosing $\kappa$ appropriately. As we are considering the bound spectrum below $E_c=-1/2$, we have in any case $E+1/2<0$. Setting thus
	$0< \kappa^{2} = |E+1/2|$, we obtain
	\begin{equation}
		- \frac{ \partial }{ \partial x } \frac{1}{m(x)} \frac{ \partial }{ \partial x } \psi_{2} + V(x) \psi_{2} = -\kappa^{4} \psi_{2},
		\label{psi2-eq}
	\end{equation}
	with
	\begin{eqnarray}
		m(x) &=& \frac{ x^{2} +1 }{ x^{2} \frac{4r}{(1+r)^{2}} + 1 },
		\label{m}\\
		V(x) &=& - \frac{1}{4} \frac{\Delta^{2} - \left( \Delta_{c}^{(1/4)} \right)^{2} }{\left( x^{2} + 1 \right)^{2} } \left( \frac{ \Delta - \Delta_{c}^{(3/4)} }{ \Delta + \Delta_{c}^{(1/4)} } x^{2} + 1 \right). \nonumber \\
		\label{V}
	\end{eqnarray}
	
	For $\Delta = \Delta_{c}^{(1/4)}$, we have $V(x) = 0$ and no bound spectrum exist as proven in section~\ref{sec-4}.
	
\begin{figure}
		\centering
		\includegraphics[width=0.9\linewidth]{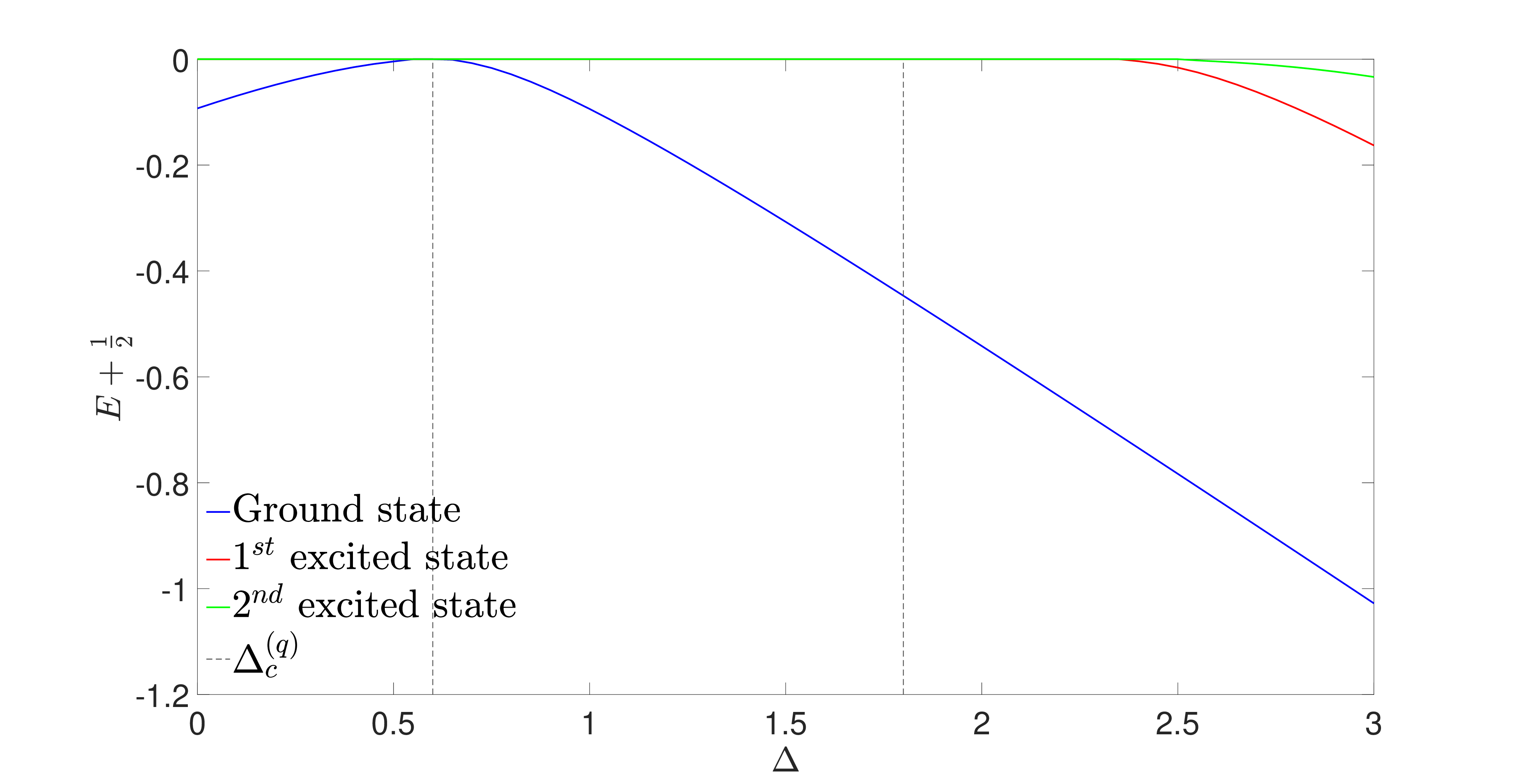}
		\caption{The ground state and the numerically resolvable excited states are shown as blue, red, and green lines for anisotropy  $r=0.25$. $\Delta_{c}^{(1/4)} = 0.60$ and $\Delta_{c}^{(3/4)} = 1.80$ have been marked by dashed lines.}
		\label{Ground_xp_fig}
	\end{figure}

	We have computed the eigenvalue $-\kappa^{4}$ for the ground state and some excited states using a spatial difference method for $0\le\Delta\le 3$. As shown in Fig.~\ref{Ground_xp_fig}, the value of $E_{GS}+1/2$ is maximal at $\Delta_{c}^{(1/4)}$ as expected, while discrete levels below $E_c$ exist for $\Delta\neq\Delta_c^{(1/4)}$. These numerically computed states are non-degenerate contrary to a claim in \cite{lo_spectral_2021} that all discrete eigenvalues are doubly degenerate. We demonstrate in appendix \ref{app-B} that all bound states are non-degenerate, while the states in the continuum above $E_c$ are four-fold degenerate due to the $\mathbb{Z}_4$-symmetry.
	
	The function $m(x)$ in \eqref{m} is larger than zero for all $x$ and approaches $\alpha^{-1}=(1+r)^2/4r<\infty$ for $|x|\rightarrow\infty$ (we exclude the RWA case $r=0$). Therefore, we may introduce the new independent variable
	\begin{equation}
		y(x)=\int_0^x \textrm{d}x' m(x')=\frac{x}{\alpha}-\frac{1-\alpha}{\alpha^{3/2}}\textrm{arctan}(\sqrt{\alpha}x),
	\end{equation}
	and write \eqref{psi2-eq} after multiplication with $1/m(x)$ as
	\begin{equation}
		-\frac{\textrm{d}^2\psi_2}{\textrm{d}^2y}+V_2(y)\psi_2=-\kappa^4\alpha\psi_2
		\label{psi2-y}
	\end{equation}
	with
	\begin{align}
		&V_2(y)=\kappa^4\frac{1-\alpha}{1+x^2} - \frac{1}{4m(x)} \times \nonumber \\
		& \left[\frac{(\Delta-\Delta_c^{(1/4)})( \Delta - \Delta_c^{(3/4)})}{(1+x^2)^2}x^2 + \frac{\Delta^2-\left( \Delta_c^{(1/4)} \right)^{2}}{(1+x^2)^2}\right],
		\label{V2}
	\end{align}
	where $x=x(y)$. The effective potential $V_2(y)$ depends on the eigenvalue $\kappa$ which multiplies the repulsive first term in \eqref{V2}. To discuss qualitatively the number of bound states, we may use Faddeev's criterion \cite{drazin1989}: The potential $V_2(y)$ allows for at most a finite number of bound states if
	\begin{equation}
		I_1=\int_{-\infty}^\infty\textrm{d}y |V^{(-)}_2(y)|(1+|y|) < \infty,
		\label{II}
	\end{equation}
	where $V_2^{(-)}(y)=0$ whenever $V_2(y)>0$ and coincides with $V_2(y)$ elsewhere.
	The question of whether there are bound states at all and if, how many of them, are determined by values of $\kappa \gtrsim 0$, so the first repulsive term in \eqref{V2} cannot change the qualitative picture deduced from the other terms. To determine the finiteness of $I_1$, only the behavior of $V_2(y)$ for large $y$ is relevant, which means $x(y)\sim \alpha y$ and $m(y)\sim \alpha^{-1}$.
	
	There are three regions in $\Delta$ for fixed $0<\alpha<1$.\\
	A)\ $\Delta<\Delta_{c}^{(1/4)}$ \\
	We have for $\kappa=0$,
	\begin{equation}
		V_2(y) \sim \frac{\gamma}{y^2} + \frac{\gamma'}{y^4}, \quad |y| \rightarrow \infty,
		\label{v-asym}
	\end{equation}
	with $\gamma < 0$ and $\gamma'>0$. The second term is asymptotically negligible compared to the first, so $I_1$ diverges logarithmically. This entails infinitely many bound states. However, these bound states are located close to the threshold, because we must have
	\begin{equation}
		\kappa^4(1-\alpha)<(4\alpha)^{-1}(\Delta_c^{(1/4)}-\Delta)(\Delta_c^{(3/4)}-\Delta).
		\label{asym}
	\end{equation}
	Thus, the ground state energy $E_{GS}$ approaches $E_c$ for $\Delta \rightarrow\Delta_c^{(1/4)}$, although there are always infinitely many states between $E_{GS}$ and $E_c$ which are difficult to resolve numerically.\\
	B)\ $\Delta_c^{(1/4)}<\Delta<\Delta_c^{(3/4)}$\\
	The asymptotic behavior of $V_2(y)$ is reversed, with $\gamma>0$ and $\gamma'<0$. However, the negative term is dominated by the positive, so $I_1$ is finite and there are at most finitely many bound states. In this case, the region with $V_2(y)<0$ is bounded. Therefore we may apply
	a criterion for compactly supported potentials to check whether bound states are present or not. If
	\begin{equation}
		I_2 = \int_{-\infty}^\infty \textrm{d}x \frac{V_2(y(x))}{m(x)}\le 0,
	\end{equation}
	there exists at least one bound state \cite{brownstein2000}. $I_2$ can be computed explicitly. It is a quadratic function in $\Delta$ and vanishes for $\kappa=0$ at $\Delta=\Delta_c^{(1/4)}$, while being negative for $\Delta>\Delta_c^{(1/4)}$. This entails that for any $\Delta>\Delta_c^{(1/4)}$, there is at least one bound state, as corroborated by the numerical results in Fig.~\ref{Ground_xp_fig}. It is possible that there is only a single bound state in this parameter region as the exceptional $G$-function $G^{(1/4)}_+(0,g)$ has no zeroes (see middle panel in Fig.~\ref{G_nond_fig}) and therefore no additional states cross the $0^{th}$ pole line. An analytical proof could be based on the monotonicity of $G^{(q)}_\pm(0,g)$ close to $g_c$ but this has not yet been done.\\
	C)\ $\Delta>\Delta_c^{(3/4)}$\\
	Here we have $\gamma,\gamma' <0$ and there exists an infinity of bound states like in case A, most of which are exponentially close to the threshold according to \eqref{asym} and the numerical results in Figs.~\ref{snd_m0_fig} and \ref{G_nond_fig}.
	
	An interesting special case is $\Delta=\Delta_c^{(3/4)}$. Then $\gamma=0$ and $\gamma'<0$, which means that the asymptotically dominant term $\sim y^{-2}$ with positive coefficient is proportional to $\kappa^4$. For any finite $\kappa$ the region with $V_2(y)<0$ does not extend to infinity, but because $I_2<0$ we have again at least one bound state.
	
	\section{Summary}
	
	In this work, we derived the $G$-function for the anisotropic two-photon Rabi model in a compact form using the Bogoliubov operator approach and $su(1,1)$ algebra. The zeros of the $G$-function determine the energy spectrum with well-defined generalized $\mathbb{Z}_4$-parity. The pole structure of the $G$-function determines the ``spectral collapse" phenomenon at the critical coupling strength $g_{c}$. The exceptional solutions, consisting of all doubly degenerate eigenvalues, are located along the pole lines of the $G$-function and the condition for their occurrence is derived analytically. This enables us to precisely identify the condition for the existence of a fully continuous spectrum without bound states.
	
	A completely continuous spectrum occurs in the atpQRM when the qubit splitting is  $\Delta_{c}^{(1/4)} = \frac{1-r}{1+r}$, suggesting that the light-matter interaction renormalizes the effective photon frequency to zero in this case.
	When $\Delta^{(1/4)}_{c} < \Delta \leq \Delta^{(3/4)}_{c}$, a finite number, probably only one bound state exists while for other qubit splittings $\Delta$ there are infinitely many bound states below the continuum threshold. These bound states are not visible using standard numerical techniques because they are located \emph{exponentially} close to the threshold energy,
	\begin{equation}
		E_c-E_n\sim e^{-cn},
	\end{equation}
	with some positive constant $c$. This behavior differs from the familiar Coulomb potential where the highly excited states approach the threshold like a power law, $-E_n\sim n^{-2}$.  To find these elusive states numerically, we have employed the exceptional $G$-function whose zeros yield the non-degenerate solutions on the lowest pole line. This function is easier computable and allows much higher truncation numbers than exact diagonalization. Our technique may provide insights into cavity QED phenomena, such as superradiance, entanglement generation between qubits, and highly populated cavity-squeezed states. The spectral collapse of the tpQRM has been demonstrated in circuit QED \cite{felicetti_two-photon_2018}, and the experimental realization of the atpQRM in circuit QED has been proposed in \cite{cui_exact_2017}. Identifying the system’s critical properties within this exact formalism holds practical significance and defines a key direction for future research.
	
	\begin{acknowledgments}
		This work is supported by the National Key R$\&$D Program of China under Grant No. 2024YFA1408900 and by the Deutsche Forschungsgemeinschaft (DFG, German Research Foundation) under Grant No. 439943572.
	\end{acknowledgments}
	
	\newcommand{\Lq}{\Lambda^{(q)}}
	\newcommand{\xq}{\xi^{(q)}}
	\newcommand{\aq}{a^{(q)}}
	\newcommand{\bq}{b^{(q)}}
	\newcommand{\cq}{c^{(q)}}
	\newcommand{\dq}{d^{(q)}}
	\newcommand{\tdq}{\tilde{d}^{(q)}}
	\newcommand{\hq}{h^{(q)}}
	\newcommand{\thq}{\tilde{h}^{(q)}}
	\appendix
	
	\section{}
	\label{app-A}
	The term $\left[ 2 \left( n+q-\frac{1}{4} \right)\right]!/(2^n n!)$ in the $G$-function grows as $n!$ which has to be compensated by rapidly decreasing coefficients $e_n^{(q)}, f_n^{(q)}$ from \eqref{recur_tpQRM}.
	
	Small numerical errors in the recursive computation of  $e_n^{(q)}, f_n^{(q)}$ amplify upon multiplication with the factor $\left[ 2 \left( n+q-\frac{1}{4} \right)\right]!/(2^n n!)$ and lead to erroneous results especially close to $g=g_c$.
	A more stable numerical procedure consists of defining
	\begin{eqnarray}
		\Lambda_{n}^{(q)} = \frac{\left[ 2 \left( n+q-\frac{1}{4} \right)\right]!}{2^{n} n!} e_{n}^{(q)} \tanh^{n}{\theta}, \nonumber \\
		\xi_{n}^{(q)} = \frac{\left[ 2 \left( n+q-\frac{1}{4} \right)\right]!}{2^{n} n!} f_{n}^{(q)} \tanh^{n}{\theta},
	\end{eqnarray}
	so that the $G$-function reads
	\begin{equation}
		G_{\pm}^{(q)} = \sum_{n=0}^{ +\infty }  \Lambda_{n}^{(q)} \pm \xi_{n}^{(q)} .
		\label{gfunc}
	\end{equation}
	The new variables $\Lambda_n^{(q)}$ and $\xi_n^{(q)}$ satisfy the recurrence relations
	\begin{subequations}
		\begin{eqnarray}
			\Lambda_{n}^{(q)} &=& \frac{\left[ \frac{\Delta}{2} - 2g^{2}(1-r^{2}) (n+q) \right] \xi_{n}^{(q)}}{2(n+q) \beta_{+} \beta_{-} - \frac{1}{2} - E} \nonumber \\
			&& + \frac{g (1-r)} {2 \sqrt{r}} \frac{ (1+r) \beta_{-} \xi_{n-1}^{(q)} - (1-r) \beta_{+} \Lambda_{n-1}^{(q)} }{2(n+q) \beta_{+} \beta_{-} - \frac{1}{2} - E} \nonumber \\
			&& \times \frac{2 \left( n+q-\frac{1}{4} \right) \left( n+q-\frac{3}{4} \right)}{n} \tanh{\theta},	\label{rec_1}
		\end{eqnarray}
		\begin{eqnarray}
			\xi_{n+1}^{(q)} &=& \frac{ \left[-\frac{\Delta}{2} - 2g^{2} (1-r^{2}) (n+q) \right] \beta_{+} \Lambda_{n}^{(q)} }{ 4\sqrt{r}g (n+1) } \tanh{\theta} \nonumber \\
			&& + \frac{ \left[ 2(n+q) \beta_{-} (2-\beta_{+}^{2}) - \left( \frac{1}{2} + E \right) \beta_{+} \right] \xi_{n}^{(q)} }{ 4\sqrt{r}g (n+1) } \tanh{\theta} \label{rec_2}
			\nonumber \\
			&& + (1+r) \frac{ (1-r) \beta_{+} \beta_{-} \Lambda_{n-1}^{(q)} - (1+r) \beta_{-}^{2} \xi_{n-1}^{(q)} }{ 4r n(n+1) } \nonumber \\
			&& \times \left( n+q-\frac{1}{4} \right)\left( n+q-\frac{3}{4} \right) \tanh^{2}{\theta}
		\end{eqnarray}
	\end{subequations}
	
\begin{figure}[tbp]
		\includegraphics[width=1.0\linewidth]{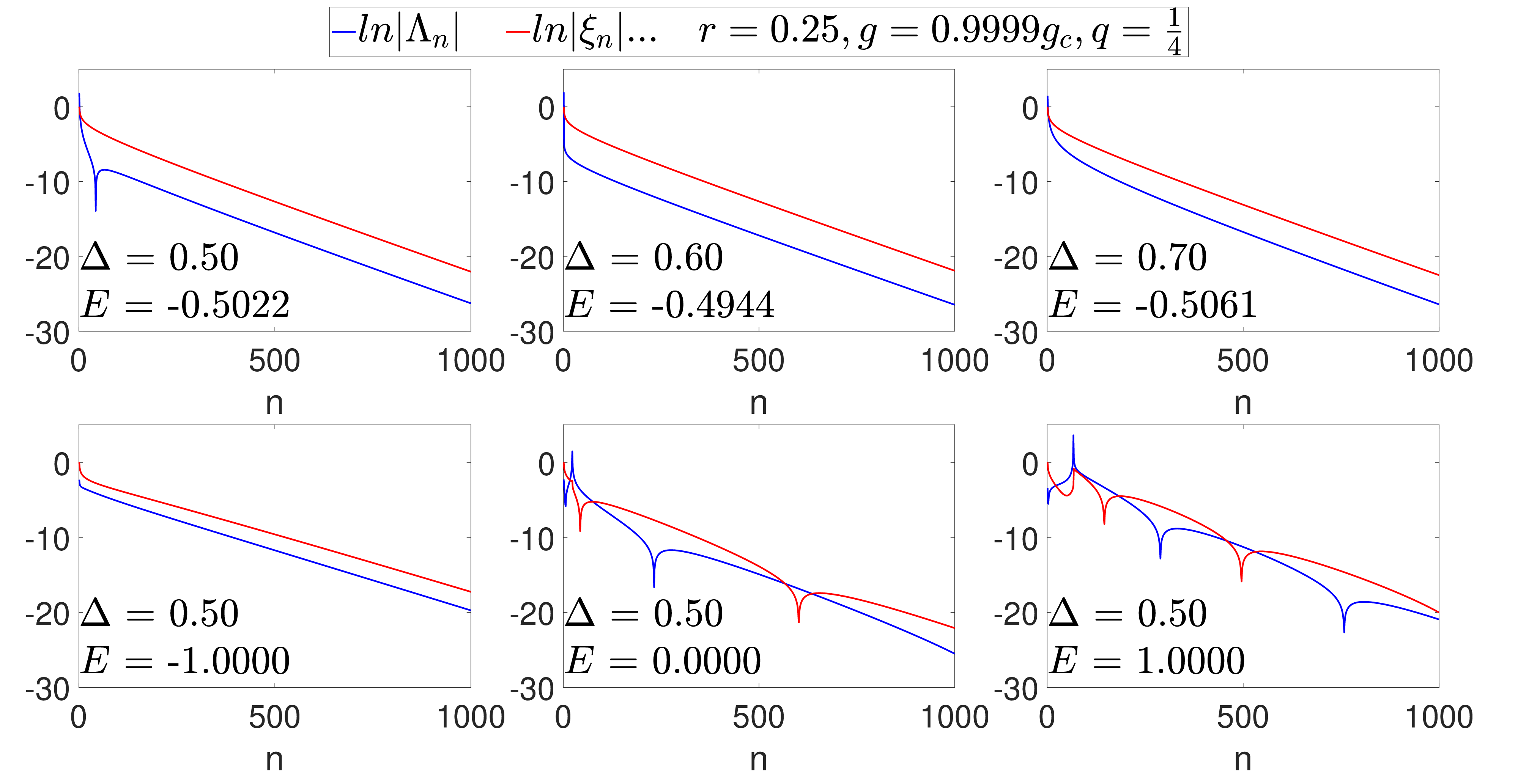}
		\caption{The upper panel gives the logarithm of parameters $\vert \Lambda_{n} \vert, \vert \xi_{n} \vert$ as $n$ increases for $E$ is the energy of the ground state, $r=0.25$, $q=\frac{1}{4}$, and $g = 0.9999 g_{c}$, corresponding to $\Delta = 0.50$ (left), $\Delta = 0.60$ (middle) and$ \Delta = 0.70$ (right). In the lower panel, $\Delta$ is fixed to $0.50$ and values of $E$ are not the eigenvalues, where $E=-1.00,0.00$ and $1.00$. Blue (red) lines mark $\Lambda_{n}$ ($\xi_{n}$).}
		\label{par_G}
	\end{figure}

	We write the coupled recurrence in a compact form
	\begin{subequations}
		\begin{align}
			\Lq_n &= \aq_n\xq_n +\bq_{n-1}\xq_{n-1}+\cq_{n-1}\Lq_{n-1}   \\
			\xq_n &= \dq_{n-1}\xq_{n-1}+\tdq_{n-1}\Lq_{n-1}+\hq_{n-2}\xq_{n-2}
			+\thq_{n-2}\Lq_{n-2}.
		\end{align}
	\end{subequations}
	The denominator $D^{(q)}_n$ of coefficients $\aq_n$ reads
	\begin{equation}
		D^{(q)}_n=2(n+q)\beta_+\beta_- - (E+1/2),
		\label{denominator}
	\end{equation}
	which contains the term $n\beta_+$. If one sets $g=g_c$ corresponding to $\beta_+=0$ in the recurrence, the denominator is $D^{(q)}_n=-(E+1/2)$ and
	\begin{equation}
		\aq_n(g_c) = \frac{2g_c(1-r^2)(n+q)-\Delta/2}{E+1/2}
	\end{equation}
	which grows linearly in $n$ and would lead to
	\begin{equation}
		\Lq_{n} \sim n!\quad \textrm{for}\quad n \rightarrow \infty
	\end{equation}
	and in turn to a diverging series expansion for the $G$-function \eqref{gfunc} with zero radius of convergence. However, for any $g$ arbitrarily close to $g_c$ such that $\beta_+\neq0$, the first term in \eqref{denominator} will dominate the asymptotics of the recurrence relation so that
	\begin{equation}
		\lim_{n\rightarrow\infty} \aq_n=a_\infty=-\frac{g^2(1-r^2)}{\beta_+\beta_-}
	\end{equation}
	stays finite. The asymptotic values of the other coefficients are
	\begin{subequations}
		\begin{align}
			b_\infty &= \frac{g(1-r^2)\tanh{\theta}}{2\sqrt{r}\beta_+},\\
			c_\infty &= -\frac{g(1-r^2)\tanh{\theta}}{2\sqrt{r}\beta_-},\\
			d_\infty &= \frac{\beta_-(2-\beta_+^2)\tanh{\theta}}{2\sqrt{r}g},\\
			\tilde{d}_\infty &= -\frac{g^2(1-r^2)\beta_+\tanh{\theta}}{2\sqrt{r}g},\\
			h_\infty &= -\frac{(1+r)^2\beta_-^2\tanh^2{\theta}}{4r},\\
			\tilde{h}_\infty &=\frac{(1-r^2)\beta_+\beta_-\tanh^2{\theta}}{4r}.
		\end{align}
	\end{subequations}
	Note that they are independent of the Bargmann index $q$. For $g<g_c$ all asymptotic coefficients are finite which means that the series
	\begin{equation}
		G^{(q)}_\pm(x)=\sum_{n=0}^\infty\left(\Lq_n+\xq_n\right)x^n
		\label{g-series}
	\end{equation}
	has a non-zero radius of convergence $R$ and exponential behavior of $\Lq_n,\xq_n$. A Poincar\'e analysis of \eqref{rec_1},\eqref{rec_2}, shows that $R>1$ which means that \eqref{gfunc} is well defined for $g<g_c$. However, the recurrence will yield uncontrolled results as long as the truncation order $N<N^*$ with the integer $N^* \sim (E+1/2)/(2\beta_+\beta_-)$. The plots in Fig.~\ref{par_G} show essentially the asymptotic region with linear behavior of $\ln\Lq_n$ and $\ln\xq_n$, because for the given parameters, $N^*$ is rather small. For example, in the lower left plot with $E<-1/2$, we have $N^*\sim 22$ and the calculation of $G^{(q)}_\pm$ will be reliable to obtain the bound states. On the other hand, the behavior of  $\ln\Lq_n$, $\ln\xq_n$ is not yet asymptotic for $E$ larger than the threshold, because $E$ lies within the quasi-continuum where $G^{(q)}_\pm$ varies strongly due to densely distributed poles. The upper panels of Fig.~\ref{par_G} show that for a value of $E$ in the discrete spectrum ($G^{(q)}_\pm=0$), the parameters become asymptotic for rather small values of $n$.
	
	\section{}
	\label{app-B}
	The discrete spectrum of the atpQRM at $g=g_c$ is non-degenerate, contrary to the statement in \cite{lo_spectral_2021}. Let's assume a normalizable eigenstate of the system \eqref{coll-H} with energy $E$ reads $\vert \Psi_+ \rangle = (\psi_{1} (x),\psi_{2}(x))^T$.
	Furthermore, we assume it to have positive parity,
	$\Pi_p|\Psi_+\rangle=|\Psi_+\rangle$.
	Then
	$|\Psi_-\rangle=( -\psi_{1} (x), \psi_{2}(x))^T$ is an eigenfunction of $\Pi_{p}$ with negative parity. In \cite{lo_spectral_2021}, it is claimed that $|\Psi_-\rangle$ is also an eigenstate of \eqref{coll-H} energetically degenerate with $|\Psi_+\rangle$.
	If so, the linear combination
	\begin{equation}
		\frac{1}{2} \left(\left(
		\psi_{1}(x), \psi_{2}(x)\right)^{T} +
		(- \psi_{1}(x), \psi_{2}(x))^{T}\right) =
		(0,\psi_{2}(x))^{T}
	\end{equation}
	would also be a solution. This entails
	\begin{eqnarray}
		\left( - \frac{\Delta}{2} + \frac{1-r}{1+r} \left( x \frac{ \partial }{ \partial x } + \frac{1}{2} \right) \right) \psi_{2} = 0 \nonumber \\ \label{eq1} \\
		- \frac{ \partial^{2} }{ \partial x^{2} } \psi_{2} = \left( E+\frac{1}{2} \right) \psi_{2}.  \nonumber \\ \label{eq2}
	\end{eqnarray}
	To make Eq.~\eqref{eq1} and \eqref{eq2} hold simultaneously, it follows that $E=-1/2$ and $\Delta = (1-r) / (1+r)$, which means only the threshold energy $E_c=-1/2$ may be doubly degenerate in this sense and then only for $\Delta_c^{(1/4)}$. These states are thus just \eqref{0-states} in section~\ref{sec-4}.

	\section{}
	
\begin{figure}[tbp]
		\includegraphics[width=1.0\linewidth]{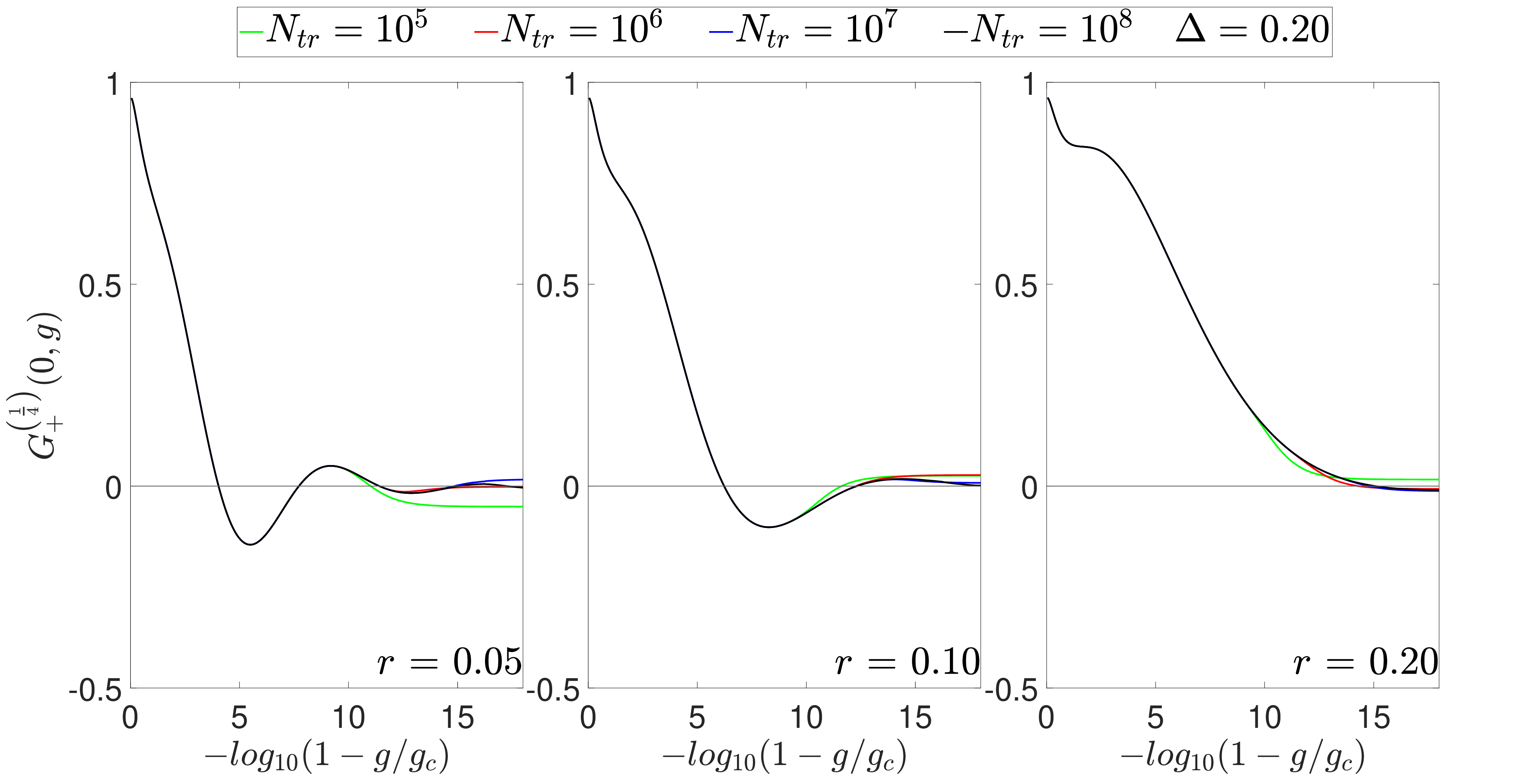}
		\caption{$G_+^{(1/4)}(0,g)$ at $\Delta=0.20$ and $r=0.05$ (left), $r =0.10$ (middle) and $r=0.20$ (right). The $x$-axis is scaled logarithmically, $x=-\log_{10} (1-g/g_{c})$. The green, red, blue, and black lines indicate that the series in \eqref{exc-gfunc} is truncated at $N_{tr} = 10^{5}$, $10^{6}$, $10^{7}$, $10^{8}$, respectively.}
		\label{non_degenerate_Delta020}
	\end{figure}

	We present results for small anisotropy $r$ in Fig.~\ref{non_degenerate_Delta020}. The number of visible bound states increases as $\Delta$ becomes smaller compared to $\Delta_{c}^{(1/4)}$, because $|\gamma|$ in \eqref{v-asym} becomes larger.
	
\begin{figure}[tbp]
		\includegraphics[width=1.0\linewidth]{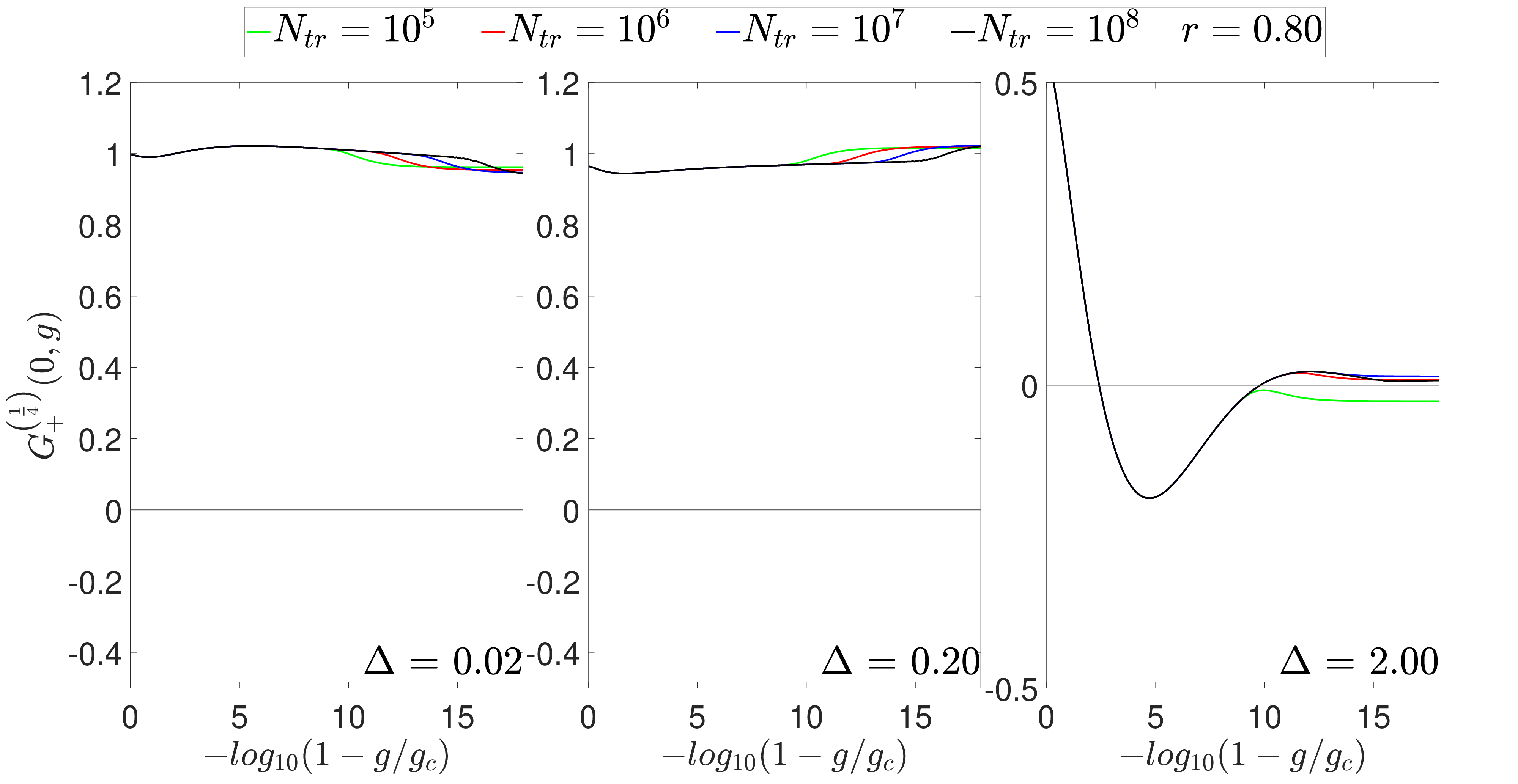}
		\caption{$G_+^{(1/4)}(0,g)$ at $r=0.80$ and $\Delta=0.02$ (left), $\Delta=0.20$ (middle) and $\Delta=2.00$ (right). The $x$-axis is scaled logarithmically, $x=-\log_{10} (1-g/g_{c})$. The green, red, blue, and black lines indicate that the series in \eqref{exc-gfunc} is truncated at  $N_{tr} = 10^{5}, 10^{6}, 10^{7}, 10^{8}$, respectively.}
		\label{non_degenerate_r080}
	\end{figure}

	Figure.~\ref{non_degenerate_r080} shows results for large $r=0.8$. The three cases A, B and C discussed in section~\ref{sec-4-1} are shown in the left (A), middle (B) and right (C) panels. The exceptional $G$-function exhibits zeros in the region $(1-g/g_c)>10^{-15}$ for case C, while it has no zeros in case B, as expected. From Figs.~\ref{G_nond_fig} and \ref{non_degenerate_r080}, we confirm that when $\Delta_{c}^{(1/4)} < \Delta < \Delta_{c}^{(3/4)}$, there is only one bound state. Case A seems to be not much different from B, as the $G$-function shows no zeroes, thus a single bound state below the threshold. In case A, the trend of the $G$-curves is to approach the zero line, so the possibility of it having a zero before $g_{c}$ cannot be ruled out, while in case B the $G$-curves trend away from the zero line and will not have any zeros. This means that in case A the additional bound states are located far in the asymptotic region. Indeed, the integral $I_2$ becomes positive for $\Delta<\Delta_c^{(1/4)}$, so the bounding effect of the potential for large $y$ requires the wave function $\psi_2(y)$ to be localized far from the center at $y=0$, with energy extremely close to $E_c$.
	
\begin{figure}[tbp]
		\includegraphics[width=1.0\linewidth]{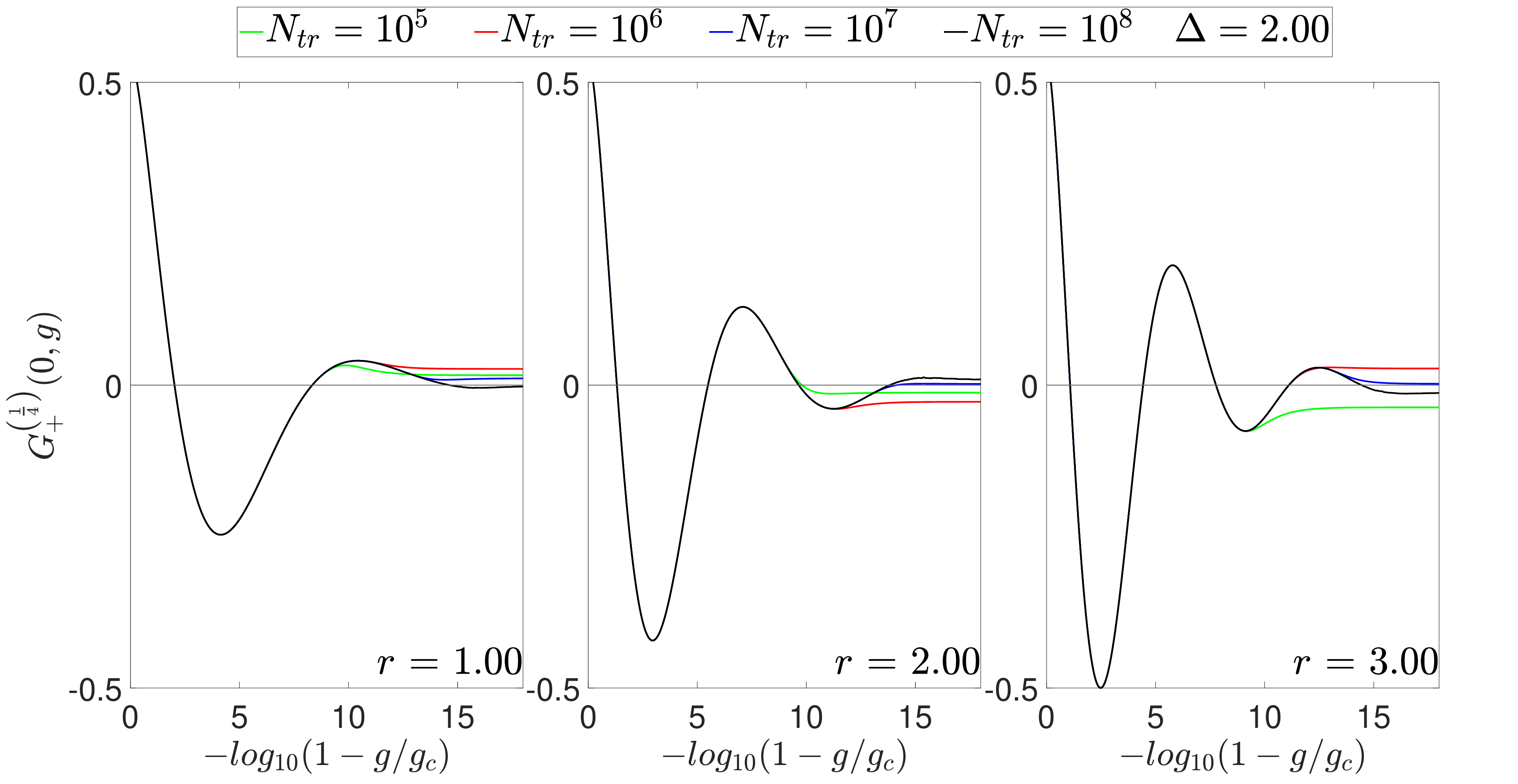}
		\caption{$G_+^{(1/4)}(0,g)$ at $\Delta=2.00$ and $r=1.00$ (left), $r =2.00$ (middle) and $r=3.00$ (right). The $x$-axis is scaled logarithmically, $x=-\log_{10} (1-g/g_{c})$. The green, red, blue, and black lines indicate that the series in \eqref{exc-gfunc} is truncated at $N_{tr} =10^{5}$, $10^{6}$, $10^{7}$, $10^{8}$, respectively.}
		\label{non_degenerate_Delta200}
	\end{figure}

	Next, we present results for $r \geq 1$, where $\Delta_{c}^{(q)} \leq 0$ and thus only case C discussed in section~\ref{sec-4-1} applies for $\Delta>0$. As shown in Fig.~\ref{non_degenerate_Delta200}, the $G$-curves exhibit similar behavior to that in the right panel of Fig.~\ref{G_nond_fig}, as expected. The zeros of $G_{+}^{(1/4)}(0,g)$ increase as the number of truncations increases, indicating infinitely many bound states.  Furthermore, the zeros $g_m$ of $G_{\pm}^{(q)}(0,g)$ appear to be distributed with almost equal distance on the logarithmic scale, which decreases as $\Delta$ or $r$ increases. We may thus approximate $g_m\sim   g_{c} [1-\exp{(-m\mu + \mu_{0})}]$, with $m$ an integer and $\mu$ and $\mu_{0}$ constants. Therefore, $g_c-g_m\propto \exp(-\mu m)$. These zeros of $G_{+}^{(1/4)}(0,g)$ correspond by continuity to bound states below $E_c$ right at $g_c$. If the level lines $E_m(g)$ passing through the $0$-th pole line at $g_c$ become almost linear with slowly varying first derivative in a plot in unscaled units for $E$ and $g$, as the numerical calculations indicate, it follows
	\begin{equation}
		E_c - E_{m} \propto \exp(-\mu m)
	\end{equation}
	Therefore, most of the bound states are located exponentially close to the threshold.
	We have seen in section \ref{sec-4-1}, that in case C the effective potential is dominated by the $y^{-2}$-term for large $|y|$ but is finite for small $y$, which corresponds to a regularization of the pure $y^{-2}$ potential similar to the short distance cut-off studied in \cite{nguyen_numerical_2020}, where the same exponential dependence of the bound state energies on the discrete label $m$ is found, in accord with our argument.
	
	\begin{figure}[tbp]
		\includegraphics[width=1.0\linewidth]{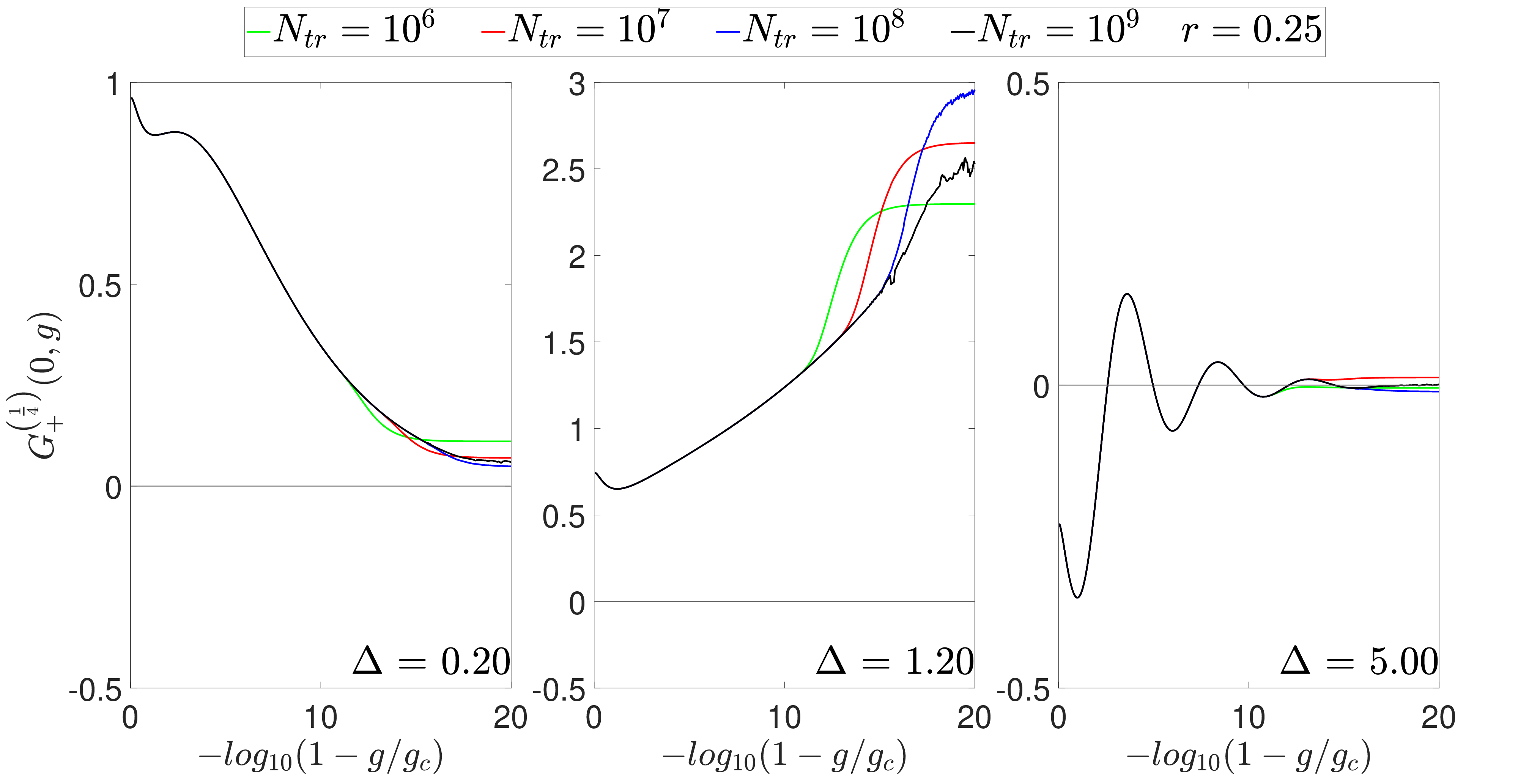}
		\caption{$G_+^{(1/4)}(0,g)$ at $r=0.25$ for the same parameters as in Fig.~\ref{G_nond_fig}, but extended up to $-log_{10}(1-g/g_c)=20$. Faint numerical noise due to limited floating point precision is visible especially in the middle panel for $1-g/g_c\le 10^{-16}$.}
		\label{non_degenerate_Ntr9}
	\end{figure}
	
	Finally, let's briefly mention the numerical challenge to compute  $G_{\pm}^{(q)}(0,g)$ exponentially close to $g_c$. Since double precision variables provide $15$ significant digits, reliable results are only obtained up to $1-g/g_{c} \sim 10^{-16}$, because the relevant variable is $\beta_{\pm} = \sqrt{1 - g^{2} (1 \pm r)^{2}}$.
	As shown in Fig.~\ref{non_degenerate_Ntr9}, numerical noise becomes visible around $1-g/g_{c} < 10^{-16}$. This problem could be remedied by applying arbitrary precision arithmetic as provided e.g. by the package GiNaC, see \cite{wolf2012}. Apart from this problem, the series expansion of  the exceptional $G$-function $G_{\pm}^{(q)}(0,g)$ is numerically remarkably stable because its convergence is that of a geometric series and it does not have poles as the regular $G$-function. However, just like the geometric series $y=(1+x)^{-1}$ requires large truncation numbers to reproduce the value $(2-\epsilon)^{-1}$ for $x=1-\epsilon$ and $0<\epsilon\approx 0$, also the evaluation of $G_{\pm}^{(q)}(0,g)$ close to $g_c$ requires an increasing number of terms because $R\rightarrow 1_+$ for $g\rightarrow {g_c}_-$ in \eqref{g-series}.

	\bibliographystyle{apsrev4-2}
	%
\end{document}